\documentclass[twocolumn,twocolappendix]{aastex63}

\def\Mgii{Mg\,{\sc ii}}

\def\Cii{C\,{\sc ii}}

\hypersetup{linkcolor=red,citecolor=blue,filecolor=cyan,urlcolor=magenta}

\usepackage{braket}
\usepackage{amsmath}
\usepackage{bm}
\usepackage{graphicx}

\received{}
\revised{}
\accepted{}

\shorttitle{The IGM Optical Depth from $z > 6.3$ Quasar Sample}
\shortauthors{Yang et al.}

\begin{document}

\title{Measurements of the $z \sim 6$ Intergalactic Medium Optical Depth and Transmission Spikes Using a New $z > 6.3$ Quasar Sample}

\correspondingauthor{Jinyi Yang}
\email{jinyiyang@email.arizona.edu}

\author[0000-0001-5287-4242]{Jinyi Yang}
\altaffiliation{Strittmatter Fellow}
\affil{Steward Observatory, University of Arizona, 933 N Cherry Ave, Tucson, AZ, USA}

\author[0000-0002-7633-431X]{Feige Wang}
\altaffiliation{NHFP Hubble Fellow}
\affil{Steward Observatory, University of Arizona, 933 N Cherry Ave, Tucson, AZ, USA}
\affil{Department of Physics, University of California, Santa Barbara, CA 93106-9530, USA}

\author[0000-0003-3310-0131]{Xiaohui Fan}
\affil{Steward Observatory, University of Arizona, 933 N Cherry Ave, Tucson, AZ, USA}

\author[0000-0002-7054-4332]{Joseph F. Hennawi}
\affil{Department of Physics, University of California, Santa Barbara, CA 93106-9530, USA}

\author[0000-0003-0821-3644]{Frederick B. Davies}
\affil{Lawrence Berkeley National Laboratory, 1 Cyclotron Rd, Berkeley, CA 94720-8139, USA}

\author[0000-0002-5367-8021]{Minghao Yue}
\affil{Steward Observatory, University of Arizona, 933 N Cherry Ave, Tucson, AZ, USA}

\author[0000-0003-2895-6218]{Anna-Christina Eilers}
\altaffiliation{NHFP Hubble Fellow}
\affil{MIT Kavli Institute for Astrophysics and Space Research, 77 Massachusetts Ave., Cambridge, MA 02139}

\author[0000-0002-6822-2254]{Emanuele P. Farina}
\affil{Max Planck Institut f\"ur Astrophysik, Karl--Schwarzschild--Stra{\ss}e 1, D-85748, Garching bei M\"unchen, Germany}

\author[0000-0002-7350-6913]{Xue-Bing Wu}
\affil{Kavli Institute for Astronomy and Astrophysics, Peking University, Beijing 100871, China}
\affil{Department of Astronomy, School of Physics, Peking University, Beijing 100871, China}

\author[0000-0002-1620-0897]{Fuyan Bian}
\affil{European Southern Observatory, Alonso de C\'ordova 3107, Casilla 19001, Vitacura, Santiago 19, Chile}

\author[0000-0001-9879-7780]{Fabio Pacucci}
\altaffiliation{BHI \& Clay Fellow}
\affil{Black Hole Initiative, Harvard University, Cambridge, MA 02138, USA}
\affil{Center for Astrophysics $\vert$ Harvard \& Smithsonian, Cambridge, MA 02138, USA}

\author[0000-0001-9299-5719]{Khee-Gan Lee}
\affil{Kavli Institute for the Physics and Mathematics of the Universe (WPI), University of Tokyo, Kashiwa 277-8583, Japan}



\begin{abstract}
We report new measurements of the intergalactic medium (IGM) Ly$\alpha$ and Ly$\beta$ effective optical depth at $5.3<z<6.5$, using a new sample of quasar sightlines including 32 quasars at $6.308\le z\le7.00$. These quasars provide a large statistical sample to measure the IGM evolution during the transition phase of the reionization epoch. We construct a data set of deep optical spectra of these quasars using VLT, Keck, Gemini, LBT, and MMT. We measure the Ly$\alpha$ effective optical depth at $5.36<z<6.57$ using the Ly$\alpha$ forests of both individual spectra and the stacked spectrum. The large scatter of individual measurements is consistent with previous work, suggesting an inhomogeneous reionization process.
Combining our new measurements and previous results, we obtain a best-fit for the Ly$\alpha$ effective optical depth evolution at $z>5.3$, $\tau\propto(1+z)^{8.6\pm1.0}$.
We then estimate the observed Ly$\beta$ effective optical depth using Ly$\beta$ forests and convert them to Ly$\alpha$ optical depth for comparison, which provides additional constraints on the evolution of the IGM optical depth. The Ly$\beta$-based measurements are generally in agreement with the best-fit evolution obtained from Ly$\alpha$ forests.
Using this new sample, we identify 389 Ly$\alpha$ and 50 Ly$\beta$ transmission spikes at $5.5<z<6.3$. The upper limits of Ly$\alpha$ optical depth estimated using transmission spikes are well consistent with our best-fit evolution. The evolution in number density of these high-redshift transmission spikes suggests a rapid transition phase at the end of the reionization. Comparison of our optical depth measurements with hydrodynamical simulations indicates a IGM neutral hydrogen fraction $\langle f_{\rm HI}\rangle\gtrsim10^{-4}$ at $z=6$.
\end{abstract}



\section{Introduction} \label{sec:intro}
The era during which early generations of stars, galaxies, and active galactic nuclei (AGN) ionized the neutral hydrogen in the IGM and ended the cosmic `dark ages' is known as the Epoch of Reionization (EoR). The EoR represents a critical epoch in the cosmic history. When and how reionization occurred gives unique insight into the formation of the first galaxies and the evolution of the earliest supermassive black holes (SMBHs). Understanding reionization is also crucial for fully exploiting the cosmic microwave background (CMB) and the IGM as cosmological probes. 

Current constraints on the evolution of IGM neutral fraction ($f_{\rm HI}$) with cosmic time rest on two main pillars. The first is the CMB measurements of the electron scattering optical depth to the surface of last scattering, resulting in the most robust constraint on the timing of reionization to date, with a mid-point of reionization at $z_{\rm reion}\sim7-8$ \citep{planck18}. However, the CMB only places an integral measurement, and hence the duration of reionization, $\Delta z_{\rm reion}$, or the shape of the $f_{\rm HI}$ evolution, is still poorly constrained. The second one is Ly$\alpha$ absorption spectroscopy of the most distant quasars and galaxies. Residual HI in a photoionized IGM gives rise to the Ly$\alpha$ forest absorption observed toward background quasars. Thus absorption spectra of the highest redshift quasars provide the most direct measurements of IGM evolution in the EoR. 
The observation of complete Gunn-Peterson absorptions towards many quasar sightlines at $z > 6$, along with the steep rise with redshift of both the Ly$\alpha$ optical depth and its scatter, has led to the consensus that we are witnessing the end of reionization only at $z\sim5-6$ \citep{fan06, becker15, bosman18, eilers18}. 
The detections of Ly$\alpha$ damping wing profiles in four $z>7$ quasars suggest a neutral gas fraction $\langle f\rm_{HI}\rangle \sim0.3$ - 0.7 at $z\gtrsim7$ and provide compelling evidence of the reionization phase transition at these redshifts \citep{mortlock11, banados18, davies18b, greig17,greig19, wang20, yang20}. The observations of galaxy Ly$\alpha$ visibility at $z > 6$ have inferred the IGM neutral fractions of $\langle f\rm_{HI}\rangle=0.59^{+0.11}_{-0.15}$ at $z\sim7$ and $\langle f\rm_{HI}\rangle>0.76$ at $z\sim8$ \citep{mason18,mason19}.

One of the most striking features of the IGM near reionization is that the intergalactic Ly$\alpha$ optical depth over $5 < z < 6$ has large variations between different lines of sight \citep{songaila04, fan06, becker15,bosman18,eilers18}. The Ly$\alpha$ forest in quasar spectra shows more variation in the amount of large-scale ($\sim$50 Mpc$/h$) transmitted flux at these redshifts than can be explained by differences in the density field alone. The fluctuations of the UV background (UVB) with a spatially uniform mean free path of ionizing photons is not enough to reproduce the observed large scatter in the optical depth \citep{becker15}. 
In particular, the longest ($\sim$160 comoving Mpc) and most opaque known Ly$\alpha$ trough discovered by \citet{becker15} is difficult to reproduce in simulations \citep{chardin15,daloisio15,davies16}; however, its association with an underdensity of galaxies \citep{becker18} suggests that the variations in optical depth require either strong fluctuations in the ionizing UVB \citep{davies18a, daloisio18}, or that reionization ends as late as $z\sim5-5.5$ \citep{kulkarni19,nasir19,keating20,choudhury20}, later than previously thought. Improved statistics of Ly$\alpha$ forest transmission at $z\gtrsim5.5$, and especially $z\gtrsim6$, are required to distinguish between these models.

Recent measurements of Ly$\alpha$ effective optical depth using large samples of $z \gtrsim 6$ quasar sightlines have provided solid constraints on the optical depth evolution at $5 < z < 6$ \cite[e.g.,][]{bosman18, eilers18}. \cite{bosman18} find that neither a fluctuating UVB dominated by rare quasars nor temperature fluctuations due to patchy reionization could fully capture the observed scatter in IGM optical depth. \cite{eilers18} show significantly higher optical depths at $5.3 < z < 5.7$ than previous studies and evidence for increased scatter even at $5.0 < z < 5.5$. Their results suggest that the spread in observed $\tau_{\rm eff}$ cannot be explained by fluctuations of the underlying density field alone, and thus support an inhomogeneous reionization scenario. However, the source that best explains the increased scatter remains an open question. 
Towards higher redshift, $6 < z < 7$, direct measurement of IGM optical depth remains challenging due to the small dataset of available deep optical/near-infrared spectra of luminous quasars in this redshift range. In addition, Ly$\alpha$ transition saturates for volume-averaged neutral fractions $\langle f_{\rm HI}\rangle \gtrsim 10^{-4}$, which limits its ability to probing the redshift range of $z \gtrsim 6.2$. Ly$\beta$ forests therefore becomes a better tracer at this redshift since Ly$\beta$ transition has five times smaller oscillator strength and thus is more sensitive to IGM with higher neutral hydrogen fraction. However, to date only a handful of Ly$\beta$ measurements exist \citep[e.g.,][]{fan06, eilers19}, and they have not yet been rigorously modeled \cite[but see][]{oh05, keating19}. 
 
In this paper, we present our new measurements of the IGM effective optical depth along 32 quasar sightlines at $6.308 \le z \le 7.00$. Our sample increases the number of quasar sightlines at $z>6.3$ by a factor of five compared with previous studies \citep[e.g.][]{bosman18,eilers18}. We constrain the effective optical depth in three different ways: Ly$\alpha$ effective optical depth, Ly$\beta$ effective optical depth, and transmission spikes. Our new quasar sample is assembled from the recent wide field high redshift quasar surveys \citep{mortlock11,venemans13,venemans15,mazzucchelli17,reed17,banados16,banados18}. In particular, there are 22 quasars at $z > 6.5$, including 13 from our recent $z \sim 7$ quasar survey \citep{wang18, wang19, fan19, yang19, yang20}. 
We describe the new quasar sample and dataset in Section 2. The methods that we use to measure the Ly$\alpha$ effective optical depth are presented in Section 3. We report the optical depth measurements of Ly$\alpha$ forests in Section 4. The effective optical depth measurements from Ly$\beta$ forests is described in Section 5. In Section 6, we report our search for the high redshift Ly$\alpha$ and Ly$\beta$ transmission spikes and the constraints on the Ly$\alpha$ effective optical depth using transmission spikes. We also discuss the estimates of neutral hydrogen fraction based on our measurements and a uniform UVB model in Section 7. The summary of this work is presented in Section 8.
All results below refer to a $\Lambda$CDM cosmology with parameters $\Omega_{\Lambda}$ = 0.7, $\Omega_{m}$ = 0.3, and $h$ = 0.685.

\section{The New $z \gtrsim 6.5$ Quasar Sample}
In this section we describe the spectral dataset used for the analysis of the IGM effective optical depth. 
We introduce the construction of quasar sample and the observational properties of spectra in Section 2.1 and data reduction in Section 2.2.

\begin{figure*}
\centering 
   \epsscale{1.0}
   \plotone{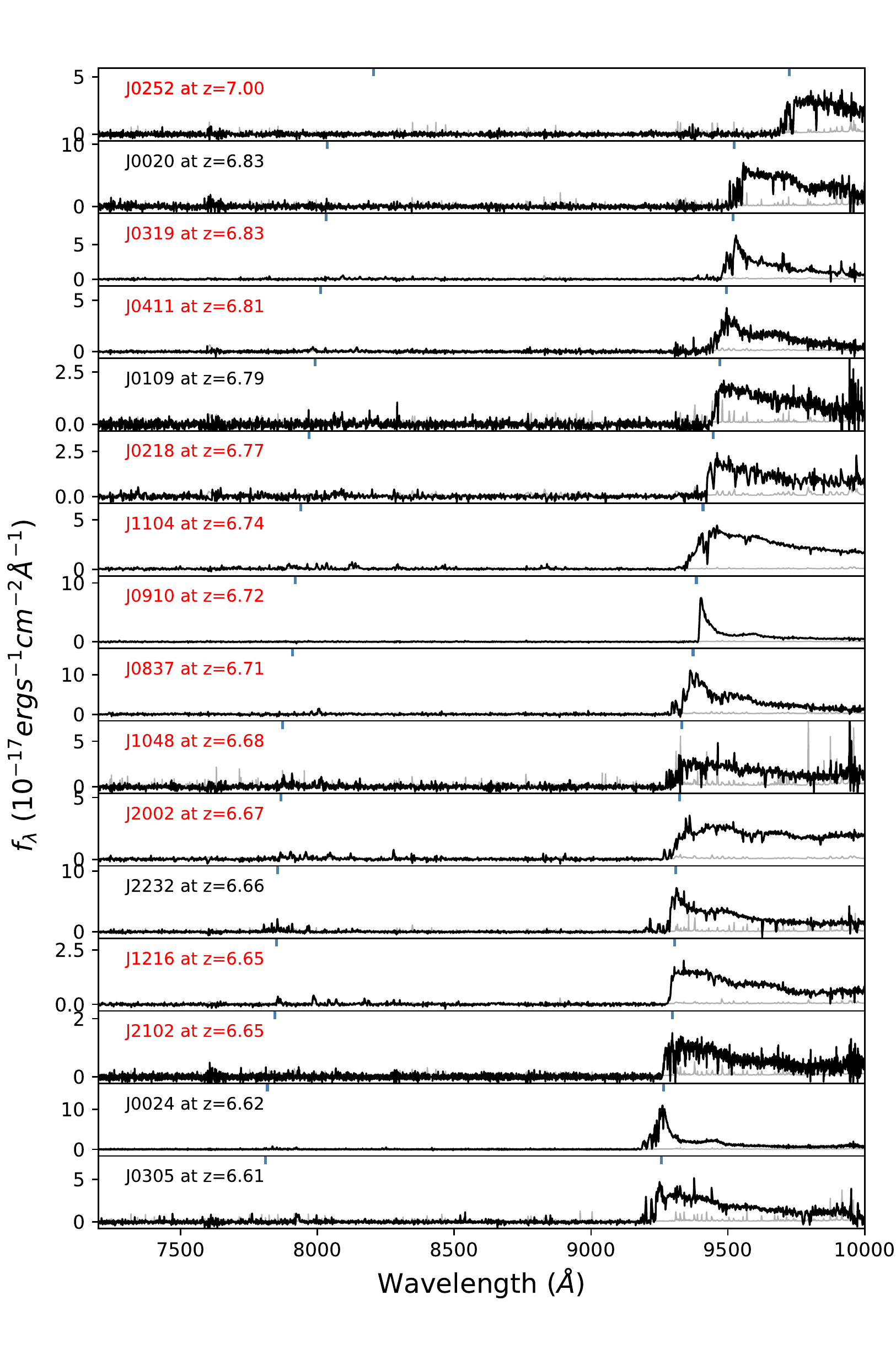} 
   \caption{Quasar spectra used for the effective optical depth measurements, sorted by redshift. The X-Shooter spectra plotted here are binned by three pixels. The grey line represents the uncertainty. The blue short lines denote the Ly$\alpha$ and Ly$\beta$ lines. Quasars with name and redshift in red are newly discovered from our survey.}
   \label{fig:spectra1}
\end{figure*}

\addtocounter{figure}{-1}
\begin{figure*}
   \centering 
   \epsscale{1.0}
   \plotone{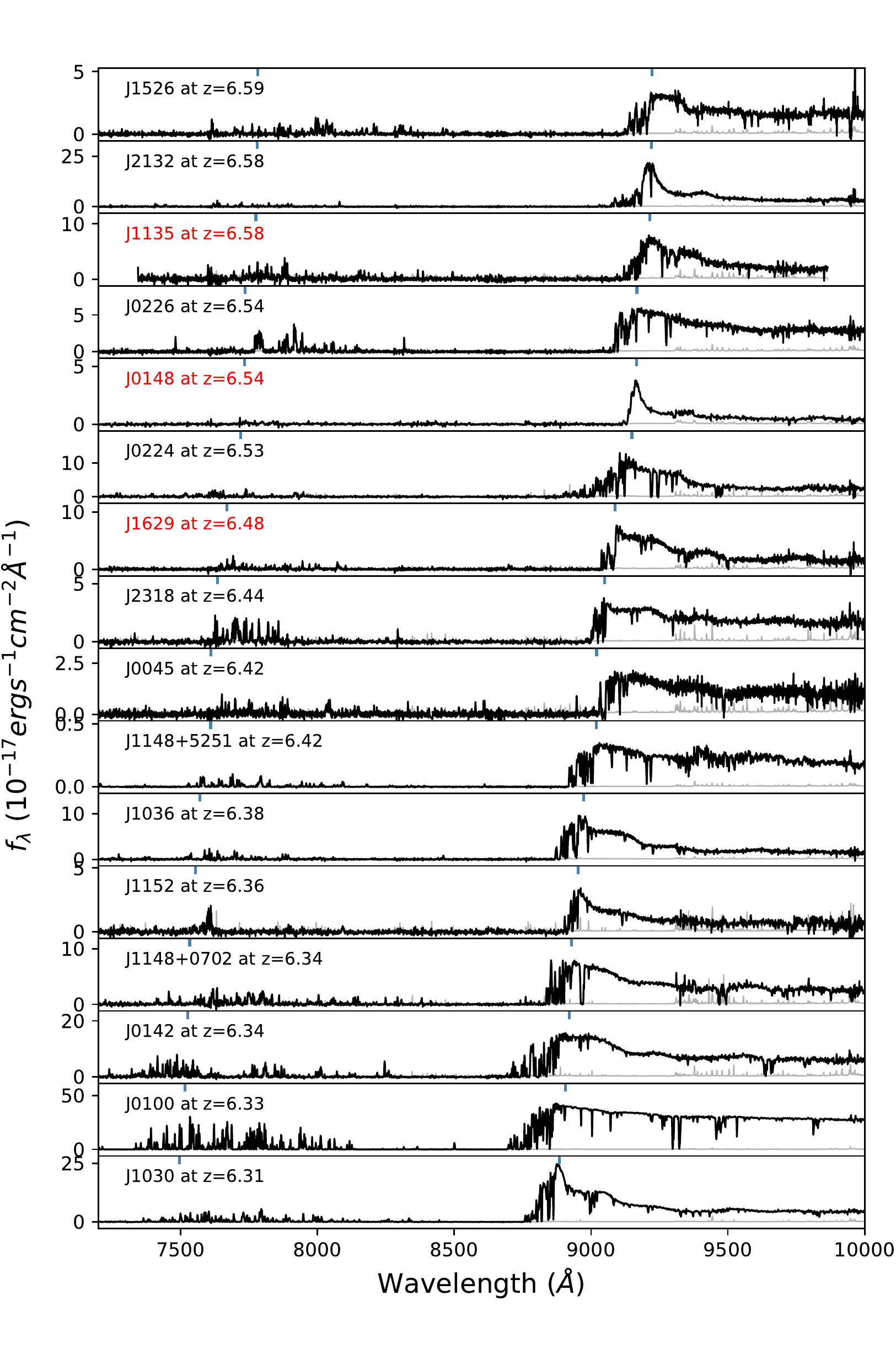} 
   \caption{Continued.}
   \label{fig:spectra2}
\end{figure*}

\begin{deluxetable*}{ l l l l l l l}
\tablecaption{Quasar Properties and Data Information of 32 Quasars in Our Data Sample.}
\tabletypesize{\scriptsize}
\tablewidth{0pt}
\tablehead{
\colhead{Name} &
\colhead{$z_{\rm em}$} &
\colhead{Instrument} &
\colhead{ExpTime(sec)} &
\colhead{S/N}\tablenotemark{a} &
\colhead{$z$ Reference\tablenotemark{b}} &
\colhead{Discovery}
}
\startdata
  J0020$-$3653 & 6.834 & VLT/X-Shooter & 4800 & 16 & Reed+2019 & Reed+2019\\
  J0024+3913 & 6.621 & Keck/DEIMOS & 10800 & 27 & Mazzucchelli+2017 & Tang+2017\\
  J0045+0901 & 6.42 & Keck/DEIMOS & 3600 & 24 & Mazzucchelli+2017 & Mazzucchelli+2017\\
  J0100+2802 & 6.327 & VLT/X-Shooter & 39600 & 445 & Wang+2019 & Wu+2015\\
  J0109$-$3047 & 6.7909 & VLT/X-Shooter & 21600 & 12 & Decarli+2018 & Venemans+2013\\
  J0142$-$3327 & 6.3379 & VLT/X-Shooter & 5580 & 34 & Decarli+2018 & Carnall+2015\\
  J0148$-$2826\tablenotemark{c} & 6.54 & Gemini/GMOS & 9600 & 11 & Yang+in prep & Yang+in prep \\
  J0218+0007\tablenotemark{c} & 6.77 & Keck/LRIS & 3600 & 8 & Yang+in prep & Yang+in prep\\
  J0224$-$4711 & 6.526 & VLT/X-Shooter & 4640 & 23 & Reed+2019 & Reed+2017\\
  J0226+0302 & 6.5412 & Keck/DEIMOS & 6000 & 52 & Decarli+2018 & Venemans+2015\\
  J0252$-$0503 & 7.00 & VLT/X-Shooter & 9600 & 9 & Wang+in prep& Yang+2019\\
  J0305$-$3150 & 6.6145 & VLT/X-Shooter & 14400 & 18 & Decarli+2018 & Venemans+2013\\
  J0319$-$1008 & 6.83 & Gemini/GMOS & 15300 & 20 & Yang+2019 & Yang+2019\\
  J0411$-$0907 & 6.81 & LBT/MODS\tablenotemark{d} & 13200 & 10 & Wang+2018 & Wang+2018\\
  J0837+4929 & 6.71 & LBT/MODS & 4080 & 13 & Wang+2018 & Wang+2018\\
    &  & MMT/BINOSPEC & 3600 & & & \\
  J0910+1656 & 6.72 & Keck/LRIS & 7200 & 19 & Wang+2018 & Wang+2018\\
  J1030+0524 & 6.308 & VLT/X-Shooter & 28300 & 118 & Decarli+2018 & Fan+2001\\
  J1036$-$0232 & 6.3809 & Keck/DEIMOS & 3600 & 55 & Decarli+2018 & Banados+2016\\
  J1048$-$0109 & 6.6759 & VLT/X-Shooter & 4800 & 10 & Decarli+2018 & Wang+2017\\
  J1104+2134 & 6.74 & Keck/LRIS & 7200 & 46 & Wang+2018 & Wang+2018\\
  J1135+5011 & 6.58 & MMT/BINOSPEC & 7200 & 20 & Wang+2018 & Wang+2018\\
  J1148+0702 & 6.344 & VLT/X-Shooter & 7200 & 48 & Shen+2019 & Jiang+2016\\
  J1148+5251 & 6.4189 & Keck/ESI & 90600 & 122 & Decarli+2018 & Fan+2003\\
  J1152+0055 & 6.3643 & VLT/X-Shooter & 31885 & 18 & Decarli+2018 & Matsuoka+2016\\
  J1216+4519 & 6.654 & LBT/MODS& 20400 & 21 & Wang+2018 & Wang+2018\\
   & & Keck/LRIS& 4800 &  &  & \\
   & & Gemini/GMOS & 4800 &  & & \\
  J1526$-$2050 & 6.5864 & Keck/DEIMOS & 7200 & 41 & Decarli+2018 & Mazzucchelli+2017\\
  J1629+2407 & 6.476 & Keck/DEIMOS & 3600 & 29 & Mazzucchelli+2017 & Mazzucchelli+2017, Wang+2018\\
  J2002$-$3013\tablenotemark{c}  & 6.67 & Gemini/GMOS & 8400 & 31 & Yang+ in prep & Yang+ in prep\\
  J2102$-$1458 & 6.648 & Keck/DEIMOS & 10800 & 15 & Wang+2018 & Wang+2018\\
  J2132+1217 & 6.585 & Keck/DEIMOS & 3600 & 44 & Decarli+2018 & Mazzucchelli+2017\\
  J2232+2930 & 6.658 & VLT/X-Shooter & 14400 & 26 & Decarli+2018 & Venemans+2015\\
  J2318$-$3113 & 6.4435 & VLT/X-Shooter & 9600 & 22 & Decarli+2018 & Decarli+2018\\
 \enddata
 \tablenotetext{a}{The average signal-to-noise ratio (S/N) per 60 km $s^{-1}$ is estimated in the rest frame wavelength range of 1245 -- 1265 \AA\ for $z < 6.75$ quasars and 1245 -- 1255 \AA\ for $z >= 6.75$ quasars. Since these quasar spectra are from several different instruments, we estimated the S/N at each pixel and convert it to S/N at 60 km $s^{-1}$ for comparison, with S/N$_{\rm 60}$ = S/N$_{\Delta v}$ $\times \sqrt{60/\Delta v}$.}
 \tablenotetext{b}{The reference of the redshifts used for optical depth measurements.}
 \tablenotetext{c}{Quasar J0148, J0218, and J2002 are unpublished new quasars discovered by our group.}
 \tablenotetext{d}{The exposure time with LBT/MODS are the sum of exposures from MOD1 and MOD2.}
 \label{tab:sample}
\end{deluxetable*}

\subsection{Dataset Construction}
The number of $z>6.5$ quasars has dramatically increased in the past two years and offers an excellent opportunity to study the IGM evolution in the reionization epoch. Our new quasar sample is based on these recent discoveries of high-redshift quasars. 
The quasar sample used for the optical depth measurements includes 32 quasars in the redshift range of $6.308 \le z \le 7.00$. Among them, there are 22 quasars at $z> 6.5$, with 13 of them from our new $z \sim 7 $ quasar survey. This survey is a new wide-area ($\sim$20,000 deg$^2$ ) systematic search for reionzation-era quasars by combining newly available optical (e.g., Pan-STARRS, DECaLs, BASS, MzLS, and DES) and infrared (e.g., UHS, ULAS, VHS, and WISE) surveys. More than 35 new $z\gtrsim6.5$ quasars have been discovered from this survey, doubling the number of luminous quasars at $z > 6.5$, within the EoR, and forming a new large sample of quasar sightlines for the IGM investigation. More details of our survey can be found in \cite{wang19} and \cite{yang19}. The other known quasars are from previous quasar surveys \citep[e.g.,][]{fan01b,fan03, mortlock11, venemans15, wu15, mazzucchelli17}.
We exclude all broad absorption line quasars and quasar with damped Ly$\alpha$ absorber (DLA) or proximate DLA features. 
The high quality optical spectra of these 32 quasars were obtained using VLT/X-Shooter, Keck/DEIMOS, Keck/LRIS, Gemini/GMOS, LBT/MODS, and MMT/BINOSPEC.

The VLT/X-Shooter \citep{vernet11} data were obtained from ESO data archive \footnote{\url{https://archive.eso.org/eso/eso\_archive\_main.html}}. The binning, individual exposure time, and slit width vary among different programs. In this work, we only reduced the VIS arm data since we do not need the coverage from UVB and NIR arms. 
The Keck/DEIMOS \citep{faber03} data were obtained in May 26th and 27th, September 13th and 14th , 2017. 
All quasars were observed with the 830G grating (R $\sim$ 3300 under 0\farcs75 slit ) with $z < 6.4$ quasars centered at 8100 \AA\ while $z > 6.4$ quasars centered at 8400 \AA. 
We chose different slit widths for different targets depending on the actual seeing conditions. 
The Keck/LRIS \citep{oke95,rockosi10} spectra were taken in 2018 March and 2019 February using grating 600/10000 in Red with 1\farcs0 slit which has a resolution of $R \sim$ 1900.
The Gemini/GMOS \citep{hook04} data came from several programs with Gemini during the 2018A, 2019A, and 2019B semesters. For each quasar, we used two wavelength setups to cover the wavelength gaps. We observed J1216+4519 with R600 Grating GMOS-N (R $\sim$ 3700 with 0\farcs5 slit) in May 15th and 16th, 2018 with one setup centered at 8600 \AA\ and the other setup centered at 8700 \AA. 
J0319--1008, J2002$-$3013, and J0148$-$2826 were observed with R400 Grating (R $\sim$ 1900 with 0\farcs5 slit) in the Gemini Queues. 
J0319--1008 were observed with GMOS-N with the central wavelengths of 9000 \AA\ and 8850 \AA, while J2002$-$3013 and J0148$-$2826 were observed with GMOS-S centered at 8600 \AA\ and 8550 \AA. 
Two spectra were taken from LBT/MODS \citep{pogge10} in 2017 -- 2019 with 1\farcs0/1\farcs2 slit in Red grating, resulting in a resolution of $R \sim 1000 -1500$. 
There were two quasars observed with MMT/BINOSPEC \citep{fabricant19} in 2018 with 270 ($R \sim$ 1400) or 600 ($R \sim 4400$) grating and 1\farcs0 long slit. 
Standard stars were taken on each night for flux calibration.
The details of observations are listed in Table \ref{tab:sample}.

\subsection{Data Reduction}
All the data were reduced using a newly developed open source python spectroscopic data reduction pipeline \citep[{\tt PypeIt}\footnote{\url{https://zenodo.org/record/3743493}},][]{prochaska20}. The image processing (i.e. bias subtracting and flat fielding) followed standard techniques. For those detectors that have multiple amplifiers, we subtracted the overscan for each amplifier separately. The wavelength solutions were calculated in the vacuum frame. We used the sky OH lines as the reference of wavelength calibration for GMOS while using arc lamps for all other spectrographs. We subtracted the sky background from individual two-dimensional frames using an optimal $b$-spline fitting sky subtraction method as detailed in \cite{kelson03}. We then extracted one-dimensional spectra from individual exposures using optimal weighting. The sensitivity functions for all instruments were derived from standard stars which were then applied to the extracted one-dimensional spectra. 

The final spectrum of each object is a stack of all individual exposures. To avoid correlated noise, we did not interpolate the individual spectra. Instead, we first determined a common wavelength grid based on the dispersion of each instrument. The wavelength grid was sampled linearly in the velocity space for X-Shooter Echelle spectrograph and linearly in the wavelength for other long-slit spectrographs. We then used the histogram technique to determine which native pixel belongs to each specific wavelength bin. The stacked flux in each wavelength bin was then computed as the mean flux density of values from all native pixels belonging to that bin\footnote{The pixels in each spectra are assumed to be statistically independent and identically distributed.}. When computing the mean flux density, each native pixel was weighted by the average square of signal-to-noise ratio (S/N) of the exposure that contains this pixel. 
Finally, we corrected the telluric absorptions on the stacked spectrum for each quasar by fitting the quasar spectrum to the telluric model grids produced from the Line-By-Line Radiative Transfer Model \citep[{\tt LBLRTM} \footnote{\url {http://rtweb.aer.com/lblrtm.html}};][]{clough05} directly. For those objects observed with multiple spectrographs, we first determine a wavelength grid based on the telluric corrected spectrum with lowest spectral resolution and then apply the same stacking procedure as described above to derive the final stacked spectrum.
All reduced spectra are shown in Figure \ref{fig:spectra1}, and all basic properties of quasar spectra are summarized in Table \ref{tab:sample}. 

When we constructed the final spectral dataset, we limit the average S/N of continuum spectra at selected wavelength range (rest frame 1245 -- 1265 \AA\ for $z < 6.75$ quasars and 1245 -- 1255 \AA\ for $z >= 6.75$ quasars) to S/N $> 7$ per 60 km/s. We measured the S/N per pixel of each spectrum at its initial resolution and converted the estimate to the S/N$_{\rm 60}$, as shown in Table \ref{tab:sample}.

\section{Measurement of IGM Optical Depth}
In this Section we describe the methods that we use to measure the Ly$\alpha$ effective optical depth. To measure the optical depth, we need to first normalize each observed quasar spectrum to its unabsorbed continuum, which is described in Section 3.1. We describe the measurements in Section 3.2. 

\subsection{Quasar Continuum Normalization}
We first normalized the quasar spectra by fitting the continuum of each quasar to a power law with a fixed slope of $\alpha = -1.5$ ($f_{\lambda}  = \lambda ^{\alpha}$). We fixed the continuum slope since for quasars at $z> 6.5$, the optical spectra typically only covers the rest frame wavelength range of $\sim$ 1260 -- 1350 \AA\ at the redshift range of our quasar sample, which is too short of accurate continuum fitting. We choose the line-free regions at rest frame 1245.0 \AA\ -- 1285 \AA\ and 1310 \AA\ --1380 \AA. For quasars at $z > 6.7$, their spectra cannot cover the wavelength range redward of 1310 \AA, so only the former wavelength range is used for the continuum fitting. 
Quasar intrinsic UV continuum was suggested to be described as a broken power law with a break at 1000 -- 1200 \AA\ \citep{telfer02,shull12}. To cover the wavelength range of Ly$\beta$ forests, we extended the best-fit continuum redward of 1000 \AA\ to the blue side with a break power law. We adopted a break at 1000 \AA\ and a slope of $\alpha_{\lambda} = -0.59$ ($\alpha_{\nu} = -1.41$) at $\lambda < 1000 $ \AA\ from \cite{shull12}.

The uncertainties of power-law continuum fitting have been suggested to be 5\% -- 20\% by previous works \citep[e.g.,][]{fan02,bosman18,bosman20,eilers18}. 
\cite{bosman20} suggest mean biases of power-law reconstructions of --9.58\% over Ly$\alpha$ forests and --12.5\% over Ly$\beta$ forests. But they also mention that this offset is due to the power-law's inability to reproduce broad emission lines and the continuum emission in-between the emission lines is very accurately captured. In this paper, the continuum uncertainties we used only represent the uncertainties caused by the power-law slope.

\cite{fan02} suggest that at high redshift the effective optical depth measurement depends logarithmically on the exact shape of the power-law continuum and the typical error of continuum fitting caused by different power-law slopes is $\sim$ 5\% of the transmitted flux $F$ assuming $\sigma(\alpha) = 0.4$. 
Since we adopt a break power-law, we estimate the continuum fitting uncertainties from different power-law slopes by assuming a $\sigma(\alpha)$ = 0.4 \citep{fan02} at the red side of 1000 \AA\ and $\sigma(\alpha)$ = 0.21 \citep{shull12} at the blue side. The power-law continuum fitting with slopes varying within the 1$\sigma$ (2$\sigma$) range result in +5.7\%/--6.1\% (+11.7\%/--11.2\%) changes on the transmitted flux in the Ly$\alpha$ forests and +10.2\%/--9.2\% (+22.8\%/--18.8\%) changes on the transmitted flux in the Ly$\beta$ forests. This uncertainty is consistent with the estimate in \cite{fan02}. When we calculate the continuum uncertainties, we follow \cite{fan02} and include the $\sigma(\alpha)$=0.21 for wavelengths shorter than 1000 \AA. We also show the effects of different power-law slopes (within 2$\sigma$ range) on the measurements of $\tau$ in Section 4 and Section 5.
As shown in Section 4.2, the continuum fitting has a negligible influence on the results of $\tau$ compared to other factors, from the comparisons between our measurements and previous works that used different method for continuum construction \citep[e.g.,][]{eilers18}.

\subsection{Effective Optical Depth Measurements}

\begin{figure}
\centering 
\epsscale{1.2}
\plotone{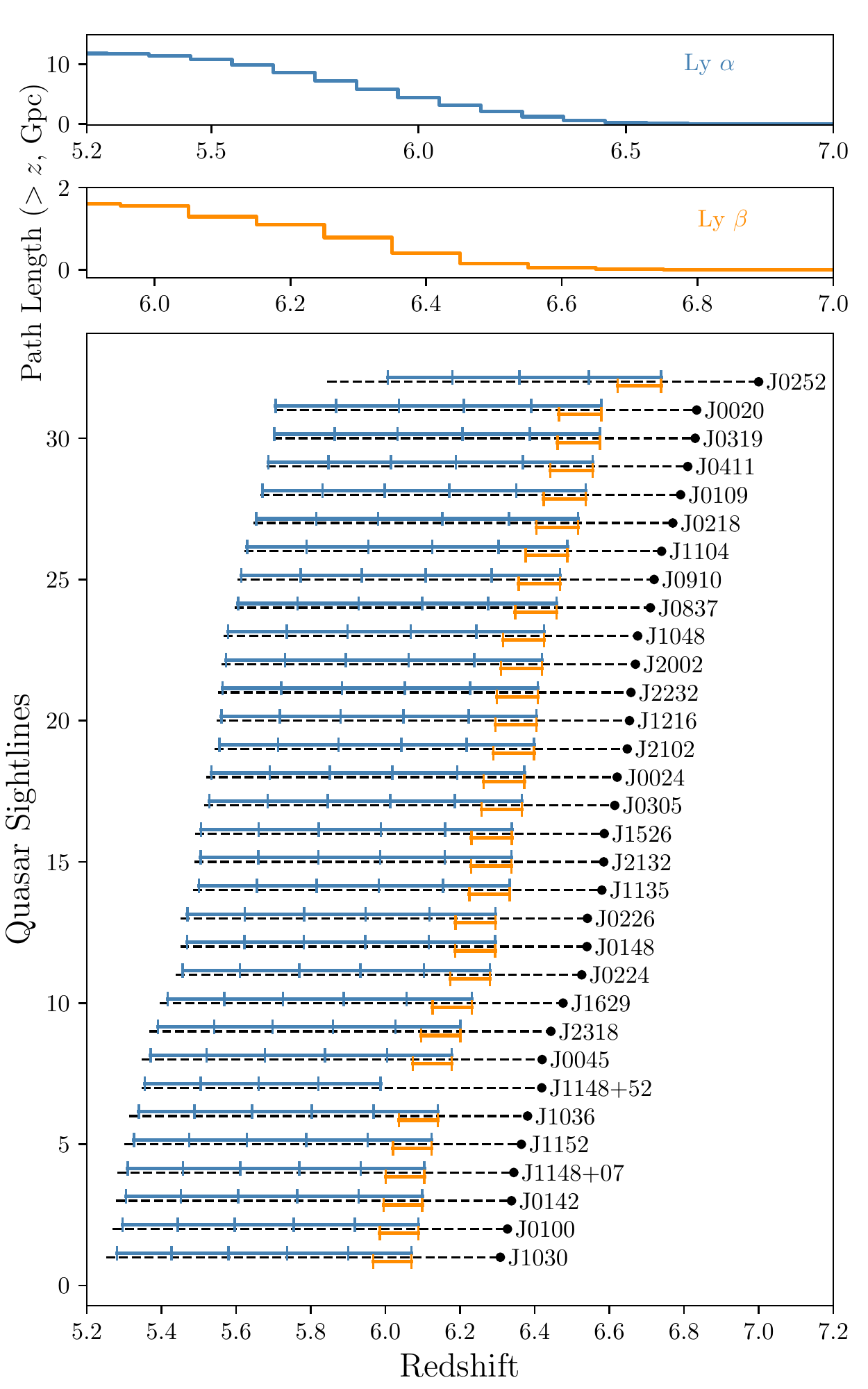} 
\caption{Redshift coverage of each quasar spectrum used for the optical depth measurements (black lines). The blue and orange lines, above and below the black lines, show the bins for Ly$\alpha$ and Ly$\beta$ effective optical depth measurements with size of 50 comoving Mpc $h^{-1}$ and 30 comoving Mpc $h^{-1}$, respectively. The wavelength windows used for Ly$\alpha$ effective optical depth measurements are limited to rest frame 1040 \AA\ -- 1176 \AA. The lower limit of wavelength window for Ly$\beta$ effective optical depth is 975 \AA\ and the upper limit is at 992.25 \AA, corresponding to the same redshift cutoff of Ly$\alpha$ forest at the high-redshift end. The unused bin of J1148+5251 is masked due to intervening low-ion absorption systems. For the same reason, there is no Ly$\beta$ measurement of J1148+5251. The two histograms represent the cumulative path length of Ly$\alpha$ and Ly$\beta$ forests in our sample at redshift higher than $z$. The total sightline length of Ly$\alpha$ is 11,839 cMpc and it is 1,604 cMpc for Ly$\beta$ sightlines.}
\label{fig:sightline}
\end{figure}

We estimate the optical depth by measuring the effective optical depth which is defined as 
\begin{equation}
\tau_{\rm eff} = -{\rm ln} \langle F \rangle
\end{equation}
where $F$ is the continuum-normalized flux. 
We measure the mean transmitted flux in fixed bins of 50 comoving Mpc $h^{-1}$ (cMpc $h^{-1}$) following \cite{becker15} and \cite{eilers18}. 
The uncertainty includes the uncertainties of spectrum and the continuum fitting. As discussed above, the 1$\sigma$ (2$\sigma$) uncertainty is about 6\% (11\%) and 10\% (20\%) in the Ly$\alpha$ and Ly$\beta$ forests, respectively.

To avoid bias caused by the quasar proximity zone, we need to choose an appropriate wavelength cut-off of Ly$\alpha$ and Ly$\beta$ forests of each quasar. Previous works chose a cut-off by measuring the quasar proximity zone size (i.e., the wavelength where the quasar's Ly$\alpha$ flux drops to 10\% of the peak value \citep{fan06, eilers18} or used a fixed value of rest frame 1176 \AA\ determined by \cite{becker15} for Ly$\alpha$ forest. 
Some quasars in our current dataset do not have \Mgii- or [\Cii]-based redshifts for the measurements of proximity zone size.
Therefore, following \cite{becker15} and \cite{bosman18}, we use rest frame 1176 \AA\ as a conservative fixed cut-off \footnote{The 1176 \AA\ cut-off means a distance of 13.5 proper Mpc from a $z=6.308$ quasar and 11.8 proper Mpc at $z=7.0$.} of Ly$\alpha$ forest. For Ly$\beta$ forest, we use the same redshift cutoff at the high-redshift end.
At the blue side, we choose rest frame 1040 \AA\ for Ly$\alpha$ forest as the minimum wavelength to avoid contamination from Ly$\beta$ or O\,{\sc vi} emissions. 
For Ly$\beta$ forest, the lower limit is set to 975 \AA\ to avoid the wavelengths affected by Ly$\gamma$ emission. 
Based on this restriction on the wavelength range, our sample including quasars at $6.308 \le z \le 7.00$ enables us to measure the effective optical depth between $5.25 < z < 6.73$, as shown in Figure \ref{fig:sightline}.

For non-detections or $< 2 \sigma$ detections of the mean transmitted flux, we adopt a lower limit on $\tau_{\rm eff}$ assuming a mean transmitted flux equal to twice the mean transmitted flux error, following previous works. 
We mask the intervening low-ion absorption systems that were identified by previous works \cite[][,Table 2]{eilers18}. We mask the entire bins including these systems (e.g., J1148+5251, J0100+2802). Specifically, these systems only affect the highest redshift bin of J1148+5251.
We do not correct for the metal line contaminations, which is also discussed by \cite{eilers19}. \cite{fauchergiguere08} show that the relative metal correction to $\tau_{\rm eff}$ in the Ly$\alpha$ forest decreases with increasing redshift from 13\% at z = 2 to 5\% at z = 4. The metal contamination at $z\sim6$ is expected to be negligible, assuming a monotonic decrease in the enrichment of the IGM over redshift \cite{eilers19}.

\section{Properties of Ly$\alpha$ Absorption}
\subsection{Ly$\alpha$ Effective Optical Depth}

\begin{figure*}
\centering 
\epsscale{0.578}
\plotone{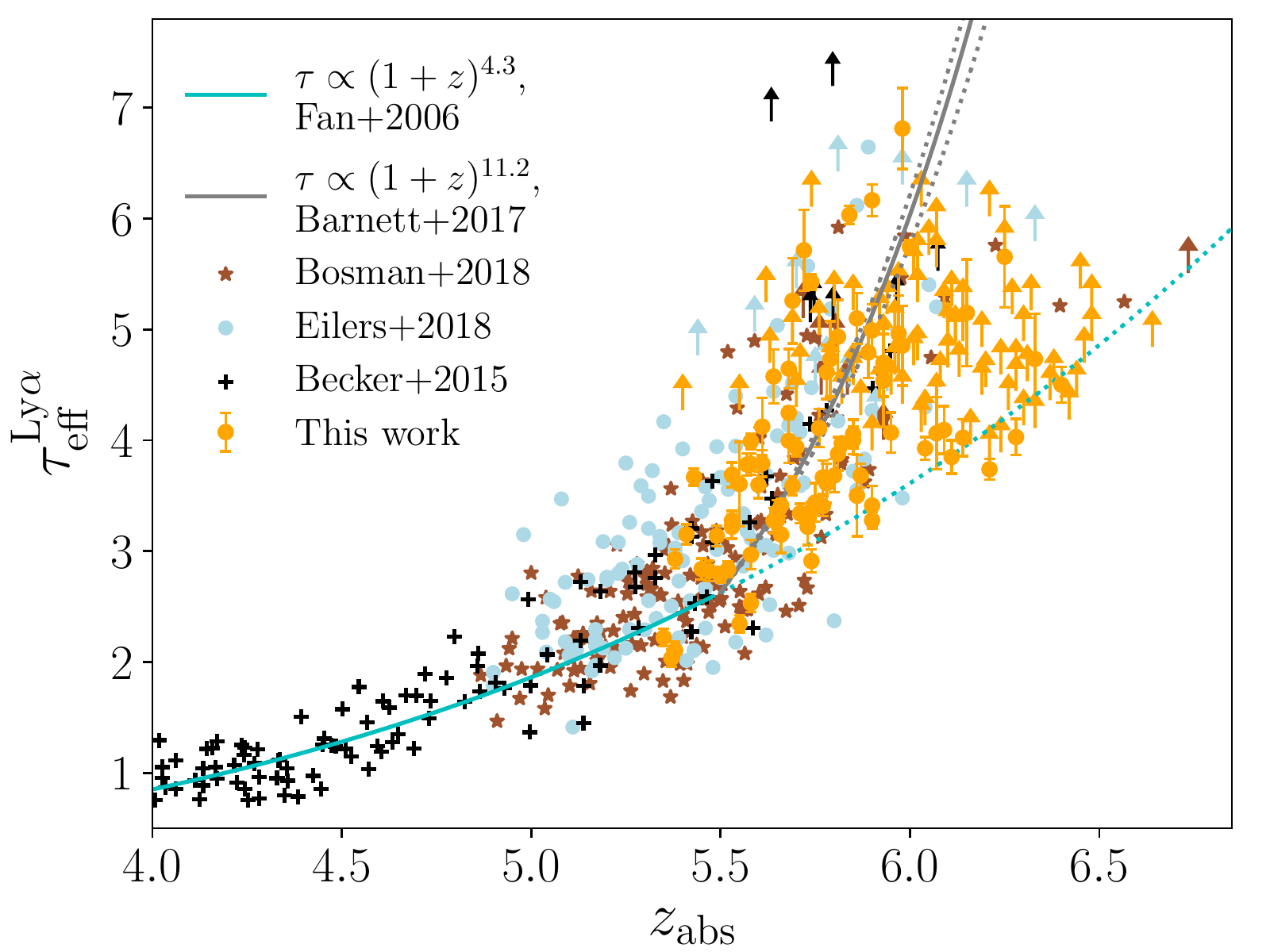} 
\plotone{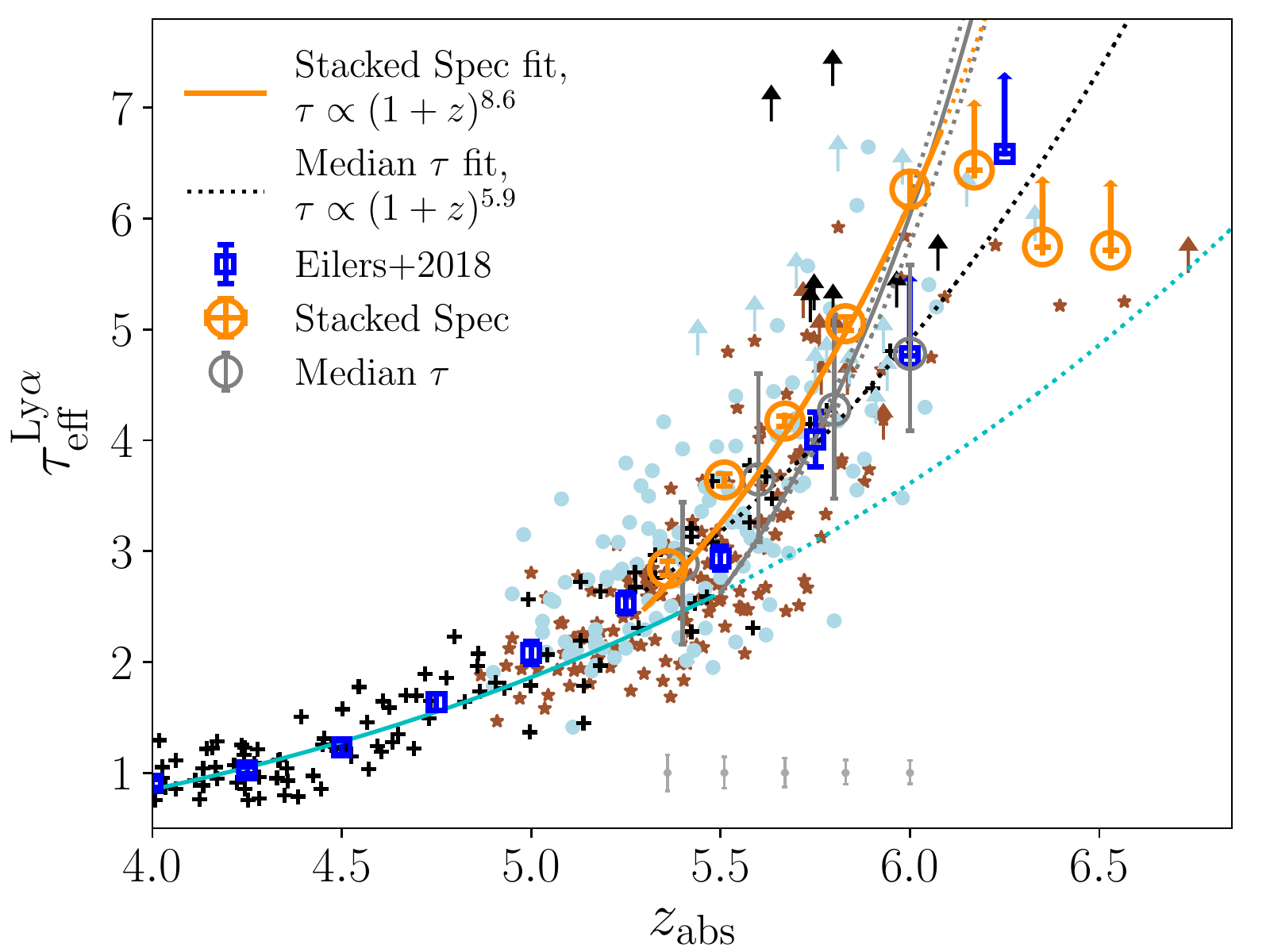} 
\caption{$Left:$ The Ly$\alpha$ effective optical depth measured from each quasar sightline in our quasar sample (light brown points/arrows), compared with previous results. 
The black crosses and the blue circles are the measurements from quasar samples in \cite{becker15} and \cite{eilers18}. The blue open squares are the mean optical depth measured in \cite{eilers18} by calculating the mean flux and the bootstrapped error on the mean in bins. The green dotted line is the effective optical depth evolution from \cite{fan06}, and the grey line represents the evolution from \cite{barnett17} with uncertainty (grey dotted lines). Our results expand the current measurements to $z \sim 6.5$, although most of the data at $z > 6.1$ are lower limits.
$Right:$ The effective optical depth measured from the stacked spectrum (orange points). The orange solid line is the best-fit of effective optical depth evolution, $\tau \propto (1+z)^{8.6\pm 1.0}$, using our measurements from the stacked spectra and the mean optical depth data from \cite{eilers18} at $z >= 5.3$ excluding all points of lower limits. The grey error bars at the bottom represent the effect of the power-law continuum fitting with different slopes (within 2$\sigma$ range) on the measurements from the stacked spectrum. We estimate this by doing the stacking and measuring the $\tau_{\rm eff}^{\rm Ly\alpha}$ using transmitted fluxes from different continuum slopes. The dark grey circles (with error bars) represent the median $\tau_{\rm eff}^{\rm Ly\alpha}$ (16th and 84th percentile) of individual measurements at redshift bins $z=$ 5.4, 5.6, 5.8, and 6.0 shown in the CDFs figure (Fig. \ref{fig:cdf}). The black dotted line shows the evolution fitting based on our median $\tau_{\rm eff}^{\rm Ly\alpha}$ and the mean optical depth data from \cite{eilers18}. The stacking of transmitted spectrum highly improves the measurement of effective optical depths at high redshift.}
\label{fig:tau}
\end{figure*}

The effective optical depths are measured from the mean transmitted flux at each bin, by using equation (1).  
All measurements are listed in Table \ref{tab:F-alpha} (Appendix A) and plotted as a function of redshift in Figure \ref{fig:tau} (left).
The spectra and the measured mean transmitted flux are also shown in Figure \ref{fig:forest1} in Appendix A.
As shown in Figure \ref{fig:tau} (left), our measurements from individual sightlines (the light brown points) span the redshift range of $5.36 < z < 6.57$. Our sample expands the current measurements of Ly$\alpha$ effective optical depth to redshift higher than 6, where there were only few previous data points. Most of points at $z > 6$ are lower limits which is due to the high neutral fraction of IGM at $z > 6$ and the limited S/N of our data. In particular, at $z \gtrsim 6.2$, the high neutral fraction will saturate Ly$\alpha$ transition. The measurements from individual sightlines are following the trend of Ly$\alpha$ effective optical depth that was estimated/predicted by previous studies. Our results show the increasing scatter of the optical depth at $5.5 \le z \le 6$, supporting the suggestion of a spatially inhomogeneous reionization reported by previous studies. 

We also stack the 1D spectra of transmitted flux in Ly$\alpha$ forests of all sightlines and measure the effective optical depth from the stacked spectrum. The stacked spectrum improves the S/N and reduce the contamination of weak absorption features or sky line residuals. 
To avoid interpolation in the stacking, we use the same stacking procedure as described in Section 2.2 but here we use an inverse-variance weighting. In our sample, there are three quasar (i.e., J0100+2902, J1030+0524, and J1148+5251) spectra that have significant higher S/N than other spectra. To avoid bias toward these three very high S/N sight lines, we lower the S/N of these three spectra to the S/N of the fourth highest S/N spectrum (i.e., S/N = 55.1) by multiplying their spectral uncertainties.
The wavelength grid of the stacked quasar transmission spectrum is generated linearly in the velocity space with a pixel size of 60 $\rm km~s^{-1}$. 

Using the stacked spectrum, we obtain the effective optical depth measurements up to $z = 6.0$ with $\tau_{\rm eff}^{\rm Ly\alpha}$ = 2.84$\pm$0.07, 3.64$\pm$0.06, 4.17$\pm$0.05, 5.05$\pm$0.06, 6.27$\pm$0.16 at $z =$ 5.36, 5.51, 5.67, 5.83, 6.0 and three lower limits, 6.44, 5.74, 5.71, at $z > 6.1$ bins (Figure \ref{fig:tau} right).
At $z>6.1$, our data cannot provide tight constraints due to the decreasing number of sightlines at high-redshift and the possible contaminations from sky lines at $\lambda > 8600$ \AA.
We then fit the evolution slope of Ly$\alpha$ effective optical depth at $5.3 < z \le 6.0$. Since our sample has a relatively small number of sightlines at the low redshift end and the stacked spectrum might be affected by high S/N sightlines, when we do the fitting we combine our new measurements at $5.36 \le z \le 6.00$ and the two mean optical depth measurements at $z > 5.3$ from \cite{eilers18} (i.e., the two measurements at $z = 5.5$ and 5.75 bins). We then obtain a new fit of the Ly$\alpha$ effective optical depth evolution over redshift at $z > 5.3$, as 
\begin{equation}
\tau \propto (1+z)^{8.6\pm 1.0}, 
\end{equation}
which is sightly above the relation from \cite{fan06} and \cite{barnett17} at $z \sim 5.3 - 5.5$ and consistent with the evolution of $\tau_{\rm eff}$ from \cite{barnett17} at $z \sim 6$ within uncertainties. 

As shown in Fig \ref{fig:tau} (right), the stacking procedure has significantly improved the measurements of effective optical depth. To check the stacking procedure and ensure that the result is not affected by any systematic effect, we test it using simulated skewers at redshift bins $z =$ 5.4, 5.6, 5.8, 6.0, and 6.2 with 2,000 skewers at each bin (see details about the simulated skewers in Section 4.2). We estimate $\tau_{\rm eff}$ from the best-fit evolution model at each redshift and assume that these values are the `true values' ($\tau_{\rm eff}$ = 2.8, 3.7, 4.8, 6.1, and 7.8). We scale the noise-free simulated skewers by matching their median $\tau_{\rm eff}$ with the `true value' at each redshift bin. We then mimic scaled skewers to match the observations based on the noise and resolution of the sightlines that are included at each redshift bin. For each skewer we randomly assign one sightline and we rebin the skewer onto the same wavelength grid of this sightline and convolve the skewer with a Gaussian that has a width equal to the resolution. After that, we add noise to skewer's transmission pixel by pixel using the transmission uncertainty of this sightline within the required redshift range. The noisy skewers have median $\tau_{\rm eff}$ of 2.8, 3.6, 4.3, 4.4, and 4.3 at different redshift bins.
 
Then we run the same stacking procedure for these noisy skewers. At each redshift bin, we randomly choose a mimicked skewer to match each observed sightline and also use the redshift of this sightline to select the skewer's transmissions within the rest frame 1040-1176 \AA. We run the randomly choosing and stacking procedure 1,001 times. From the 1,001 runs we obtain a median value of the $\tau_{\rm eff}$ measurement from the stacked transmitted spectrum at each redshift bin. We have $\tau_{\rm eff}$ = 2.8$\pm$0.1, 3.8$\pm$0.1, 4.7$\pm$0.1, and 6.0$\pm$0.2 at $z =$ 5.4, 5.6, 5.8, and 6.0 and a lower limit of 6.0 at $z = 6.2$. The agreement between the $\tau_{\rm eff}$ from the stacked spectra and the `true values' therefore supports the robustness of our spectra stacking procedure. These 1,001 runs could show the effect of cosmic variance which changes the stacked $\tau_{\rm eff}$ by $\sim$ 10\% (16th and 84th percentile) but will not change our conclusions. A larger sample of quasar sightlines will improve the measurements.

\subsection{Cumulative Distribution Functions (CDF) of Ly$\alpha$ Optical Depth}

\begin{figure*}
\centering 
\epsscale{0.9}
\plotone{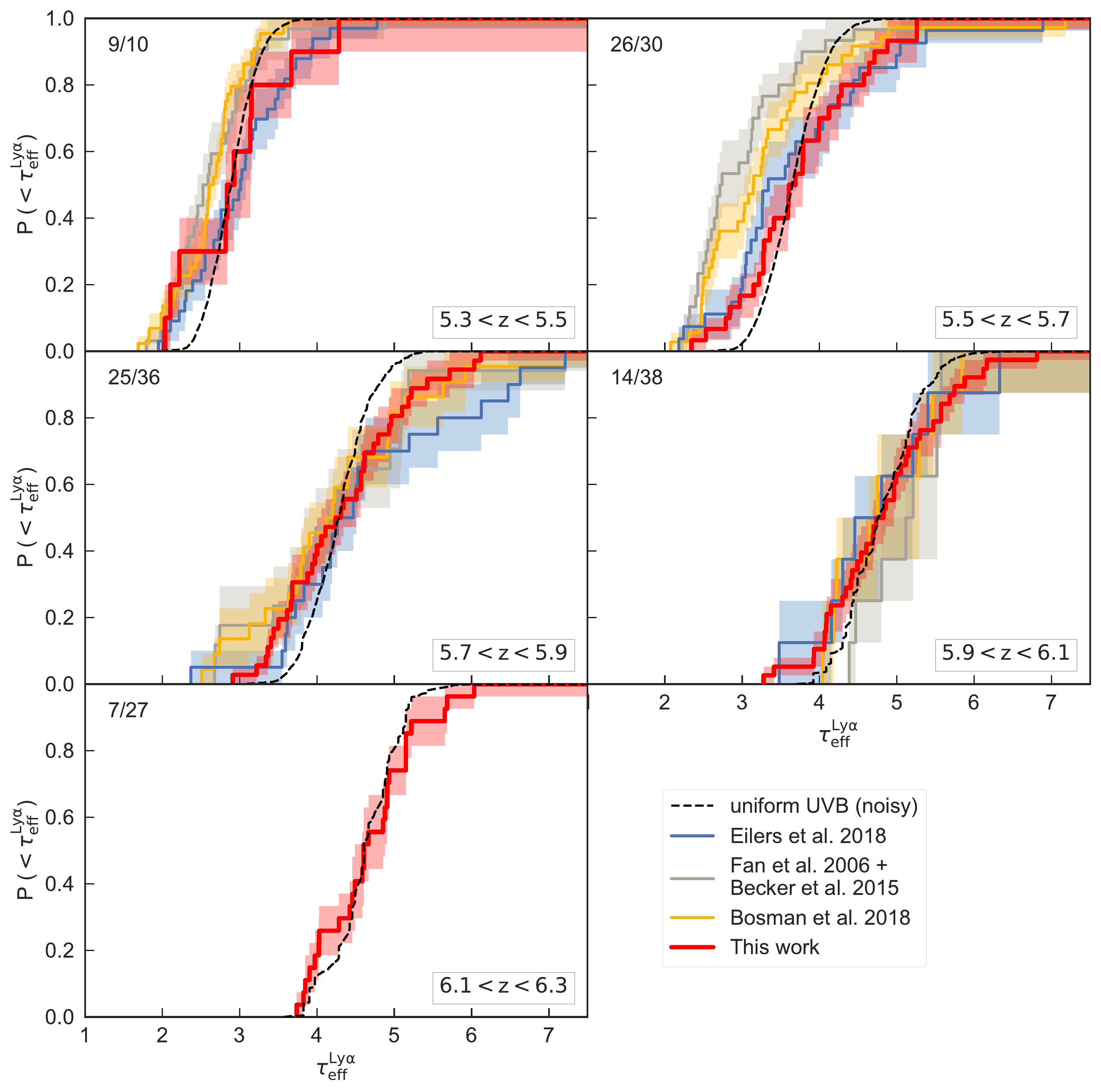} 
\caption{The CDFs of Ly$\alpha$ effective optical depth in six different redshift bins (red lines). As a comparison, we plot CDFs of previous results from \cite{fan06} and \cite{becker15} (grey), \cite{bosman18} (yellow), and \cite{eilers18} (blue). The shaded regions represent the $1 \sigma$ uncertainties estimated by bootstrapping. We treat all non-detections or $< 2\sigma$ detections of mean transmitted flux as $\tau_{\rm eff} =$ --ln($2\sigma_{\rm \langle F\rangle}$). The numbers shown at each bin (upper-left corner) indicate the fraction of $> 2\sigma$ measurements to the total measurements. In the overlapped redshift bins ($z < 6.1$), our CDFs are generally consistent with these previous results. The discrepancies could be caused by the different data quality, data reduction, and sky line masking. Our new sample provides more constraints at the $5.9 < z < 6.1$ bin than all previous samples and also enable us to explore higher redshift bins, although there the CDF is dominated by the lower limits. The black dashed lines are the CDFs from simulation skewers assuming a uniform UVB model and the skewers have been mimicked to the observations.}
\label{fig:cdf}
\end{figure*}

We calculate the CDF of the optical depths, and compare our CDF with results from previous works \citep{fan06, becker15, bosman18,eilers18} in Figure \ref{fig:cdf}. 
Following the process above and previous works, we treat all non-detections or $< 2 \sigma$ detections of mean transmitted flux as $\tau_{\rm eff} =$ --ln($2\sigma_{\rm \langle F\rangle}$). To compare the measurements from different studies, we derive the 1$\sigma$ uncertainties of all CDFs using bootstrap resampling with 5000 realizations. The result from \cite{fan06} was measured from a sample of 19 SDSS quasars at $5.74 < z_{em} < 6.42$, and the measurements in \cite{becker15} were from 42 quasars at $4.5 \le z \le 6.4$ (19 quasars are also part of the sample in \citealt{fan06}). We also plot the results from \cite{eilers18} and \cite{bosman18} (the `GOLD' sample), estimated from 23 $4.6 \lesssim z \lesssim 6.3$ quasars and 33 $z > 5.7$ quasars, respectively.

At $5.3 < z < 5.5$, our result is well consistent with \cite{eilers18} according to a two-sample KS test (p=0.9) and shows mild disagreement with the other two studies (p$<$0.3). Within the redshift bin of $5.5 < z < 5.7$ where all observations show more discrepancies between each other (p$\lesssim0.3$), our result and \cite{eilers18} are marginally consistent but have mild disagreement (p=0.3). However, the disagreements between our sample and those from \cite{fan06} + \cite{becker15} and \cite{bosman18} are more apparent (p$<$0.05). Both our result and the ones from \cite{eilers18} show systematically higher optical depths than the other two works. This discrepancy has also been discussed in \cite{eilers18} which found a systematic offset of the mean fluxes between their measurements and \cite{fan06} and the offset was stronger at lower redshifts. They suggested that the discrepancy was caused by the different data quality, continuum fitting method, data reduction pipeline, intervening absorber masking, and zero-level offsets. 
At higher redshift bins, at $5.7 < z < 5.9$ and $5.9 < z < 6.1$, our results well agree with all other samples (p$\ge$0.8) with slight discrepancies shown in the comparisons between our results with \cite{eilers18} at $5.7 < z < 5.9$ (p=0.6) and with \cite{fan06} + \cite{becker15} sample at $5.9 < z < 6.1$ (p=0.6). The former is due to the higher $\tau$ values from \cite{eilers18} at the high optical depth end and the latter is caused by our lower values at low $\tau$ end. Note that the lower limits included in observed samples could lead to uncertainties in these comparisons.

At the highest redshift bin, $6.1 < z < 6.3$, our CDF is dominated by the lower limits of $\tau$, especially at the high $\tau$ end, as shown by the significantly decreased fraction of $> 2\sigma$ measurements in Figure \ref{fig:cdf}. At this bin, there are no previous measurements, so we only compare our results with the CDF from simulation as described below. At $z > 6.3$, only two out of seventeen data points are $> 2 \sigma$ measurements. Consequently, we do not calculate the CDF at $z > 6.3$. As a conclusion, from these comparisons between our results and previous works, our result is marginally consistent with the previous observations with some mild disagreements, especially at the redshift bin of $5.5 < z < 5.7$. Compared to the results from \cite{fan06}, \cite{becker15}, and \cite{bosman18}, our result is in better statistical agreement with \cite{eilers18}. Our work is using the same continuum fitting method (i.e., fixed power law with the same $\alpha$) used by \cite{fan06}, while our result is still more consistent with \cite{eilers18} that used a different continuum modeling, which could suggest a negligible influence of the continuum fitting compared to other factors.

To compare our measurements with theoretical predictions, 
we generate simulated skewers by using a \texttt{Nyx} hydrodynamical simulations \citep{almgren13}, 100 Mpc/$h$ on a side, with 4096$^{3}$ dark matter particles and 4096$^{3}$ baryon fluid cells \citep{davies18c} and adopt the uniform UVB of \cite{haardt12}.
We extract 2000 random skewers of density, temperature, and velocity along the direction of the grid axes from simulation outputs from $z = 3.0$ to $z = 6.5$ with the step of $\delta z = 0.5$. For redshifts between the redshift bins of outputs, we take the closest output and rescale the density field by $(1+z)^{3}$. 
Then we compute the Ly$\alpha$ forest spectra in 50 cMpc $h^{-1}$ bins at $z = 5.4, 5.6, 5.8, 6.0,$ and 6.2. 
We scale the output at $z = 5.5$ for the $z = 5.4, 5.6$ bins and output at $z = 6.0$ for the next three bins.
There could be some modest systematic uncertainties because the actual structures evolve slightly between snapshots, but this uncertainty will not change the result significantly.

In order to model the influence of spectral noise and resolution on the opacity measurements, we assign resolution and noise to each skewer based on quasar sightlines at the same redshift bin. The resolution and noise assigned to a skewer come from the same sightline to fully match the observations. For each skewer, we rebin it based on the pixel size from the randomly selected sightline and convolve it with a Gaussian with width of the spectral resolution. In the high redshift Ly$\alpha$ forests where the signal from quasars are fainter than the sky background, the spectral noise is dominated by sky and read noise. So we approximately use the uncertainties of the mean transmitted flux from individual sightlines for skewers. For each measurement of mean transmitted flux from skewer, we randomly choose an error of mean transmitted flux, $\sigma_{\rm \langle F\rangle}$, from the measurements from observations at the same redshift bin. To be consistent with our real measurements, for $< 2 \sigma$ measurement of transmitted flux, we use the $\tau_{\rm eff} =$ --ln($2\sigma_{\rm \langle F\rangle}$).

Then we rescale the mimicked skewers to observations since the exact strength of the UVB radiation $\Gamma_{\rm UVB}$ is unknown in the simulation. We rescale the effective optical depth of each pixel $i$, $\tau_{i}^{\rm Ly\alpha,unscaled}$, in each skewer $j$ and then measure the mean transmitted flux of each skewer $\langle F\rangle_{j} $ and the effective optical depth $-{\rm ln}\langle F\rangle_{j}$ after scaling. We match the median $\tau$ of skewers to the median $\tau$ of our observations and compute the scaling factor $A_0$ for each redshift bin. 
\begin{align}
 \rm median  \{-{\rm ln} \langle \exp\left[-A_0 \tau_{i}^{\rm Ly\alpha, unscaled}\right]\rangle_{j}\} \nonumber\\
& = \rm median \{\tau_{\rm obs}\}
\end{align}
We obtain scaling factors of 0.66, 0.84, 0.87, 1.03, and 0.86 at the five bins, respectively. When the median $\tau$ of skewers get matched to the observations, we could see if the simulated skewers are able to predict the distribution of $\tau$ from observations under the same data conditions. As shown in Fig \ref{fig:cdf}, the comparison between the CDFs from the observed data and the simulation show the disagreement at all redshift bins except for the highest redshift bin, consistent with conclusions from previous studies, suggesting an inhomogeneous reionization. The agreement at $z=6.2$ bin does not have any physical meaning because at this bin the $\tau$ distribution is dominated by the lower limits (i.e., the uncertainties of mean transmitted flux). This also affects the comparison in the $z=6.0$ bin where we could notice a weaker discrepancy between simulation and observations then lower redshift bins.
From the four $z\le6.0$ bins, we can still notice that the rescaled uniform UVB skewers are not enough to well produce the observed distributions at both high and low $\tau$ end. The fluctuations of radiation background or/and temperature may better explain the current observations.

\section{Properties of Ly$\beta$ Absorption}

\begin{figure}
\centering 
\epsscale{1.2}
\plotone{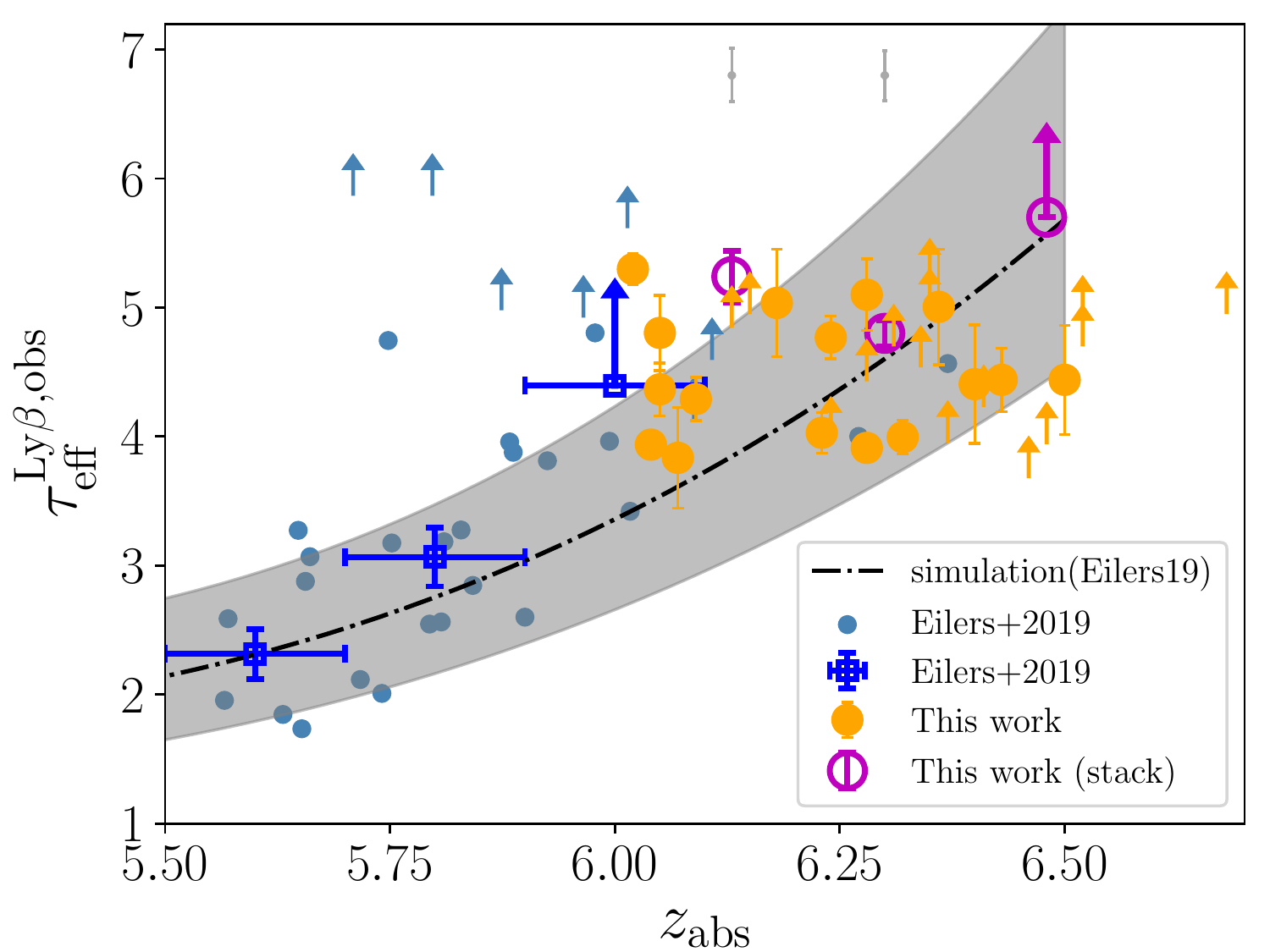} 
\caption{The observed Ly$\beta$ effective optical depths measured from individual sightlines (orange) and the stacked spectrum (purple), compared with the previous measurements from \cite{eilers19} and their Nyx hydrodynamical simulation assuming a uniform UVB (black dash-dotted line with grey shaded region). The shaded region indicating the 95th percentiles of the scatter expected from density fluctuations. The light blue points are the individual measurements and the blue squares represent the measurements averaged over redshift bins of $\Delta z = 0.2$ from \cite{eilers19}. The two grey error bars at the top represent the effect of the power-law continuum fitting with different slopes (within 2$\sigma$ range) on the measurements from the stacked spectrum.}
\label{fig:tau-beta-obs}
\end{figure}

\begin{figure*}
\centering 
\epsscale{0.57}
\plotone{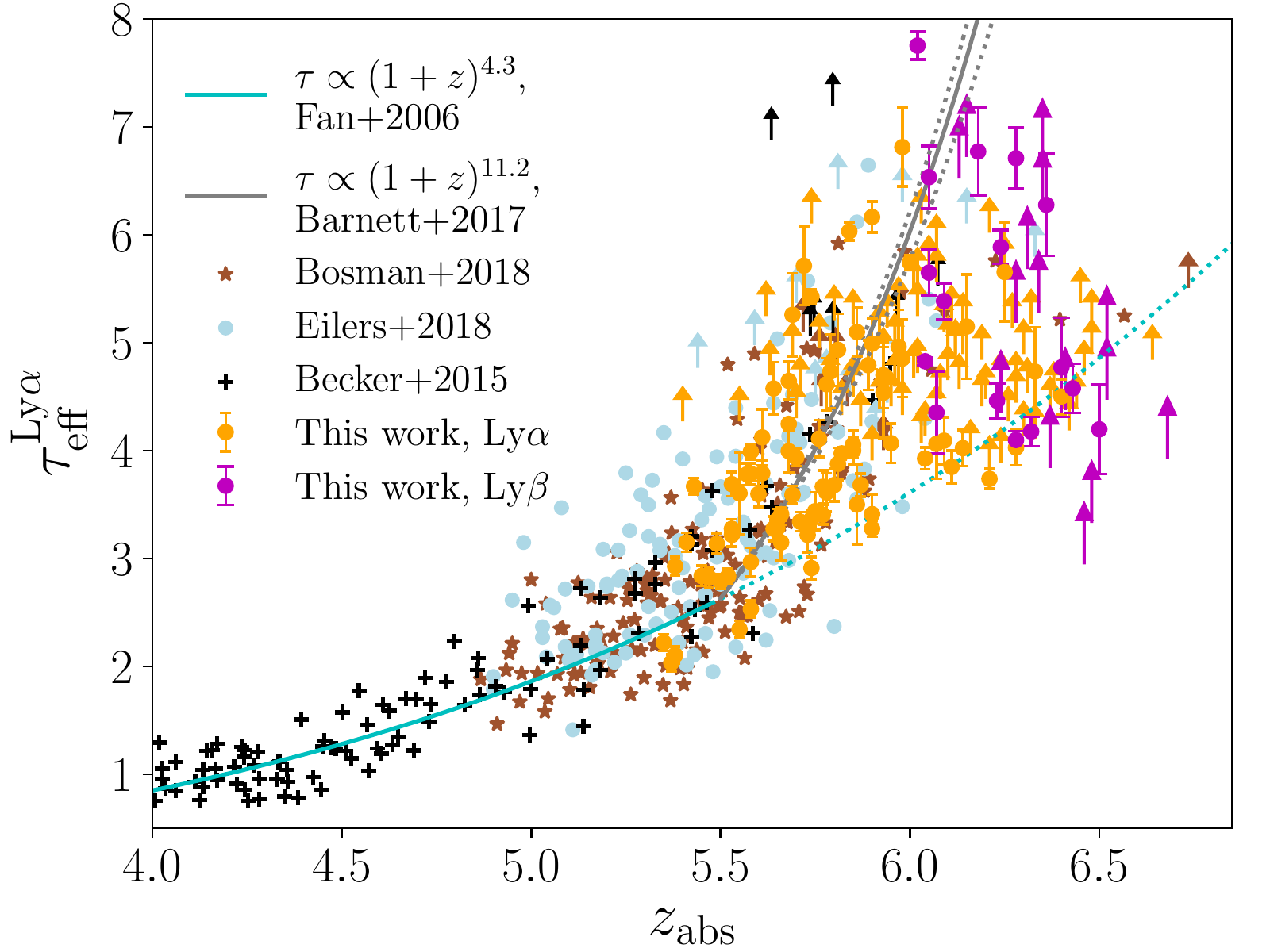} 
\plotone{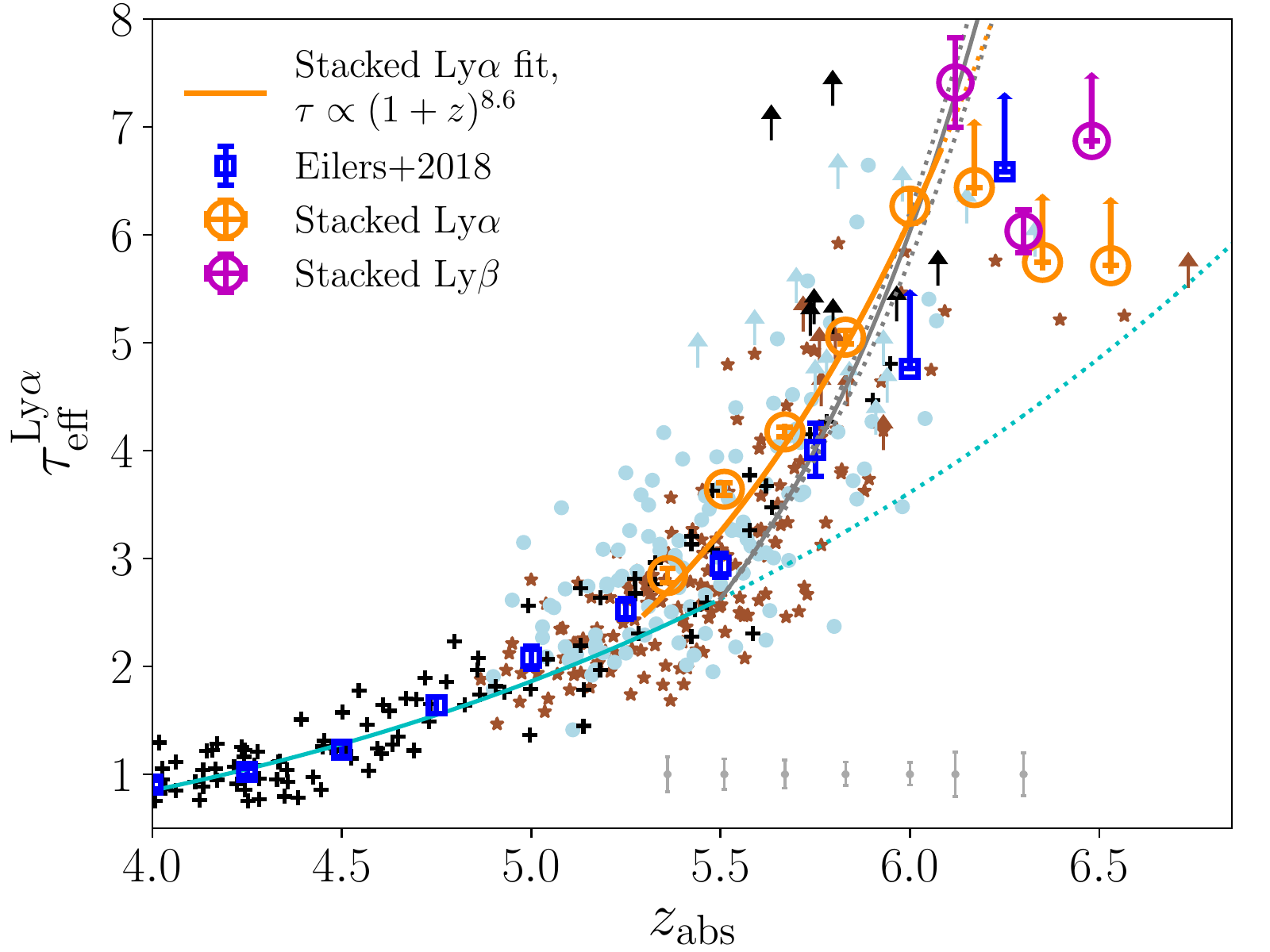} 
\caption{$Left:$ The Ly$\alpha$ effective optical depth converted from Ly$\beta$ optical depth using the conversion factor of $\tau_{\rm eff}^{\rm Ly\alpha}$/$\tau_{\rm eff}^{\rm Ly\beta}$=2.19 from \cite{fan06}. All other markers are the same to that used in Figure \ref{fig:tau}. The measurements based on Ly$\beta$ forests have similar distribution to the measurements using Ly$\beta$ forests. 
$Right:$ The effective optical depth measured from the stacked spectrum (orange points) of Ly$\beta$ forests. Due to the large uncertainties of the foreground Ly$\alpha$ effective optical depth and the conversion factors between $\tau_{\rm eff}^{\rm Ly\alpha}$ and $\tau_{\rm eff}^{\rm Ly\beta}$, the measurements converted from $\tau_{\rm eff}^{\rm Ly\beta}$ may not be used to match the Ly$\alpha$ measurements directly but could be treated as an additional constraint on the optical depth. The grey error bars at the bottom represent the effect of the power-law continuum fitting with different slopes (within 2$\sigma$ range) on the measurements from the stacked spectrum.}
\label{fig:tau-alpha-beta}
\end{figure*}

As mentioned above, compared to Ly$\alpha$, Ly$\beta$ with its $\sim 5$ times smaller oscillator strength has enhanced ability to trace neutral hydrogen at high redshift.
Therefore, we also measure the effective optical depth in the Ly$\beta$ forests. We limit the Ly$\beta$ forest in rest frame wavelength range of 975 \AA\ to 992.25 \AA. The lower limit is set to avoid the wavelengths affected by Ly$\gamma$ emission at 972.54 \AA. The upper limit is the same redshift cutoff that we applied to the Ly$\alpha$ forest to avoid proximity zone contamination, but applied to the Ly$\beta$ forest region (i.e., 1176/$\lambda_{\rm Ly\alpha}$$\times$$\lambda_{\rm Ly\beta}$ with $\lambda_{\rm Ly\alpha}$ = 1215.67 \AA\ and $\lambda_{\rm Ly\beta}$= 1025.72 \AA).

We measure the mean transmitted flux and calculate the effective optical depth using equation 1, the same as described in Section 3.2. But given the narrow redshift coverage of Ly$\beta$ forest, we use the fixed bins of 30 cMpc $h^{-1}$. We have only one bin for each sightline, as shown in Figure \ref{fig:sightline}. Since we mask the entire bin that includes the intervening absorber, we do not use the sightline of quasar J1148+5251. The Ly$\beta$ forests and mean fluxes are plotted in Figure \ref{fig:betaforest} in Appendix A. 
Note that the effective optical depth measured here is the observed Ly$\beta$ effective optical depth which includes the absorption from the foreground Ly$\alpha$ forest. 
The redshift evolution of observed effective optical depth $\tau_{\rm eff}^{\rm Ly\beta, obs}$ is plotted in Figure \ref{fig:tau-beta-obs}. Our work provides new Ly$\beta$ measurements at redshift above 6 and up to $z \sim 6.5$. Compared to previous result from \cite{eilers19}, the $\tau_{\rm eff}^{\rm Ly\beta, obs}$ is increasing rapidly but with large scatter, which is also discussed in \cite{eilers19}. A uniform UVB model can not produce the scatter as large as the observed scatter, which support the requirement of fluctuating UVB or fluctuating temperature models.  

We also measure the observed effective optical depth from the stacked transmission spectra of Ly$\beta$ forests. The stacking process for the Ly$\beta$ follows the procedure used for Ly$\alpha$ forests in Section 4.1.
Since the stacked Ly$\beta$ spectrum has long enough path, here we use fixed bins of 50 cMpc $h^{-1}$ in order to obtain higher S/N measurements.  
We exclude the sightlines of J1148+5251 and J0252--0503 for stacking. J0252--0503 is the highest redshift quasar and its Ly$\beta$ forest does not overlap with any other Ly$\beta$ forests. Therefore, the stacked spectrum Ly$\beta$ forests is based on 30 sightlines. We obtain measurements up to $z \sim 6.3$ with $\tau_{\rm eff}^{\rm Ly\beta, obs}$ = 5.24$\pm$0.19, 4.80$\pm$0.09 at $z = 6.12, 6.30$ and a lower limit (5.68) at $z= 6.48$, as shown in Figure \ref{fig:tau-beta-obs}. Combined with the result from \cite{eilers19}, we find a weak trend showing that at $z \lesssim 6.2$ the optical depth increases more rapidly than the prediction of model, which can also been seen in the individual points. However, the evolution at higher redshift is not following this trend. We could not conclude whether this trend is astrophysical or not because of the possible contamination from sky lines and small number of high quality sightlines at higher redshift ($z > 6.2$) bins. In addition, the stacked results might be affected by high S/N sightlines and the measurement at $z \sim 6.3$ could be affected by the strong transmission spikes found at $z \sim 6.2 - 6.3$ (see Section 6). Therefore, more sightlines are needed to fully characterize Ly$\beta$ absorption at this redshift range.

We then estimate the pure Ly$\beta$ effective optical depth by subtracting the foreground Ly$\alpha$ absorptions.
The Ly$\beta$ forest in our sample covers the foreground Ly$\alpha$ absorptions in redshift range of $4.8 < z < 5.6$. At $z <= 5.3$, we use the effective optical depth evolution model from \cite{fan06} to estimate the foreground Ly$\alpha$ transmitted flux. We use our new fit described in Section 4.1 for Ly$\alpha$ absorptions at $z > 5.3$. Then we compute the mean transmitted flux of pure Ly$\beta$ as $F_{\rm obs, \beta} = F_{\rm \beta} \times F_{\rm \alpha}$. To compare the pure Ly$\beta$ effective optical depth with our measurements of Ly$\alpha$, we convert the pure Ly$\beta$ optical depth to the Ly$\alpha$ effective optical depth by using the empirical conversion factor from \cite{fan06}, $\tau_{\rm eff}^{\rm Ly\alpha}$ / $\tau^{\rm eff}_{\rm \beta} = 2.19$ at $z_{abs} > 5.9$. However, the simple empirical conversion factor may not describe the relation between Ly$\alpha$ and Ly$\beta$ effective optical depth well. This value is dependent on the IGM background and higher values have also been suggested by other works. In addition, we use only the effective optical depth evolution model to estimate the foreground Ly$\alpha$ absorptions without any scatter being taken into account, which will also result in extra uncertainties. The uncertainties could also be reflected by the large scatters in Figure \ref{fig:tau-alpha-beta}. However, the individual measurements of pure $\tau_{\rm Ly\beta}$ are located in a similar region with respect to the Ly$\alpha$ measurements. Therefore, these measurements could provide additional constraints on the IGM optical depth evolution.

We also measure the pure Ly$\beta$ effective optical depth from the stacked spectrum and convert it to the Ly$\alpha$ effective optical depth, as shown in Figure \ref{fig:tau-alpha-beta}. 
The Ly$\beta$ forest window is much narrower than that of the Ly$\alpha$ forests and thus the overlapped wavelength ranges between these sightlines are smaller. Therefore, the measurements from stacked Ly$\beta$ forests spectra have larger uncertainties. The measurement at $z=6.1$, $\tau_{\rm eff}^{\rm Ly\beta}$=(3.38$\pm$0.19)$\times$2.19, is consistent with the best-fit evolution model from our Ly$\alpha$ measurements, while the higher redshift one ((2.76$\pm$0.09)$\times$2.19 at $z=6.3$) is not. This again needs to be constrained with more sightlines, as discussed above.
In addition, these Ly$\beta$-based measurements are affected by the conversion between $\tau_{\rm eff}^{\rm Ly\alpha}$ and $\tau_{\rm eff}^{\rm Ly\beta}$ and the uncertainty of the foreground Ly$\alpha$ optical depth.

\section{Transmission Spikes}
Narrow Ly$\alpha$ transmission spikes have been found up to redshifts at $z\sim6.1$ \citep[e.g.][]{chardin18} and possible Ly$\beta$ spikes have been found up to $z \sim 6.8$ \citep{barnett17}. 
The occurrence of these transmission spikes have been suggested to correspond to the most underdense regions of the intergalactic medium to be already nearly fully ionized at those redshifts \citep{chardin18, kakiichi18, garaldi19}. 
The distribution of the transmission spikes is sensitive to the exact timing and the topology of reionization. 
We search for the transmission spikes in all sightlines in our sample and use them to constrain the IGM evolution. The redshift range of our sightline sample allows the search for high redshift transmission spikes up to $z \sim 6.7$, although the search is limited by the spectral resolution and data quality.
We only search for Ly$\alpha$ and Ly$\beta$ transmission spikes at $z > 5.5$.
To be consistent, all spikes at rest frame wavelength redward of 1176 \AA\ are conservatively treated as transmissions in quasar proximity zones, although some of them may not be affected by the quasar proximity zone. Further search for spikes at redder wavelength will be carried out after we combine this optical sample with our NIR spectral dataset and measure the quasar proximity zone sizes. 

\begin{figure}
\centering 
\epsscale{1.2}
\plotone{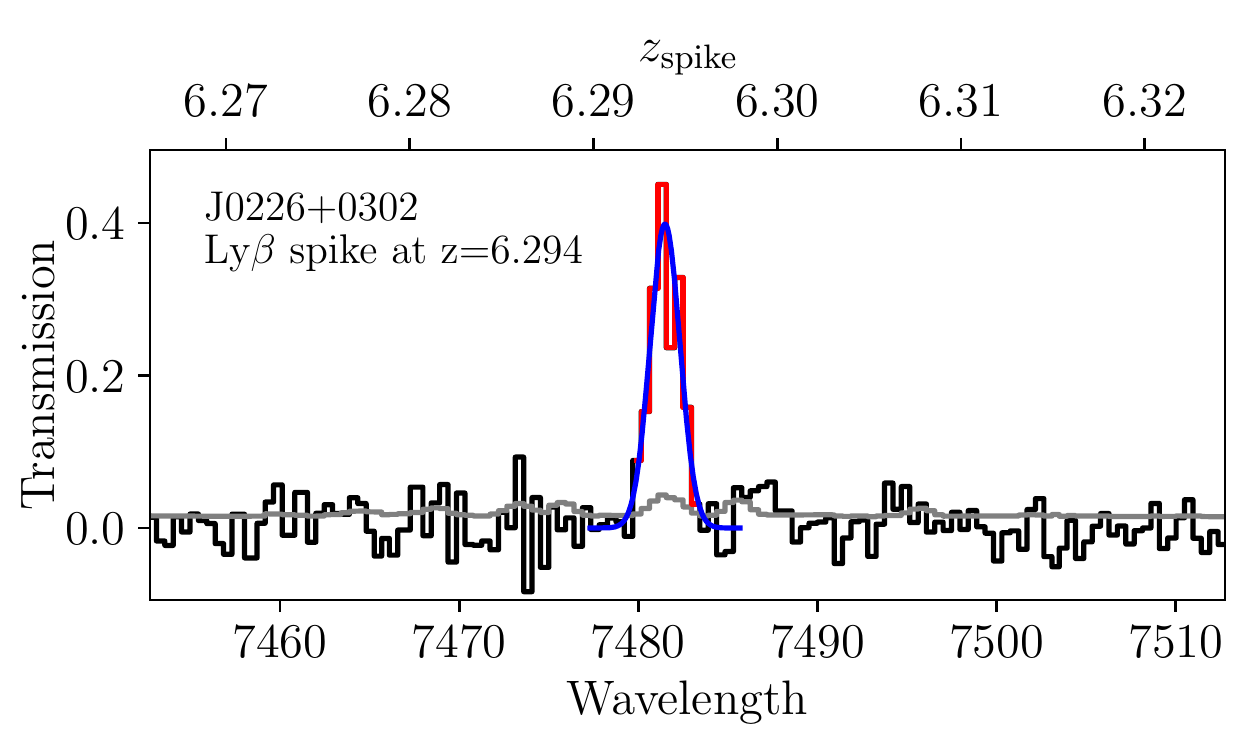} 
\caption{An example illustrating how transmission spike strength is measured. The spike plotted here is a Ly$\beta$ transmission spike identified at $z=6.294$ in the spectrum of quasar J0226+0302. The black line is the transmission spectrum and the grey line shows the uncertainty. We used two different ways to quantify spike strength, measuring 1) strength $W$ by direct integration of continuum normalized transmitted flux over the wavelength range of a spike (red) and 2) EW from the best-fit profile via Gaussian fitting (blue). The $W$ and EW obtained of this spike are consistent, with value of 0.11$\pm$0.01. All measurements of other spikes are listed in Table \ref{tab:spike-alpha} and \ref{tab:spike-beta}.}
\label{fig:spike-strength}
\end{figure}

\begin{figure}
\centering 
\epsscale{1.15}
\plotone{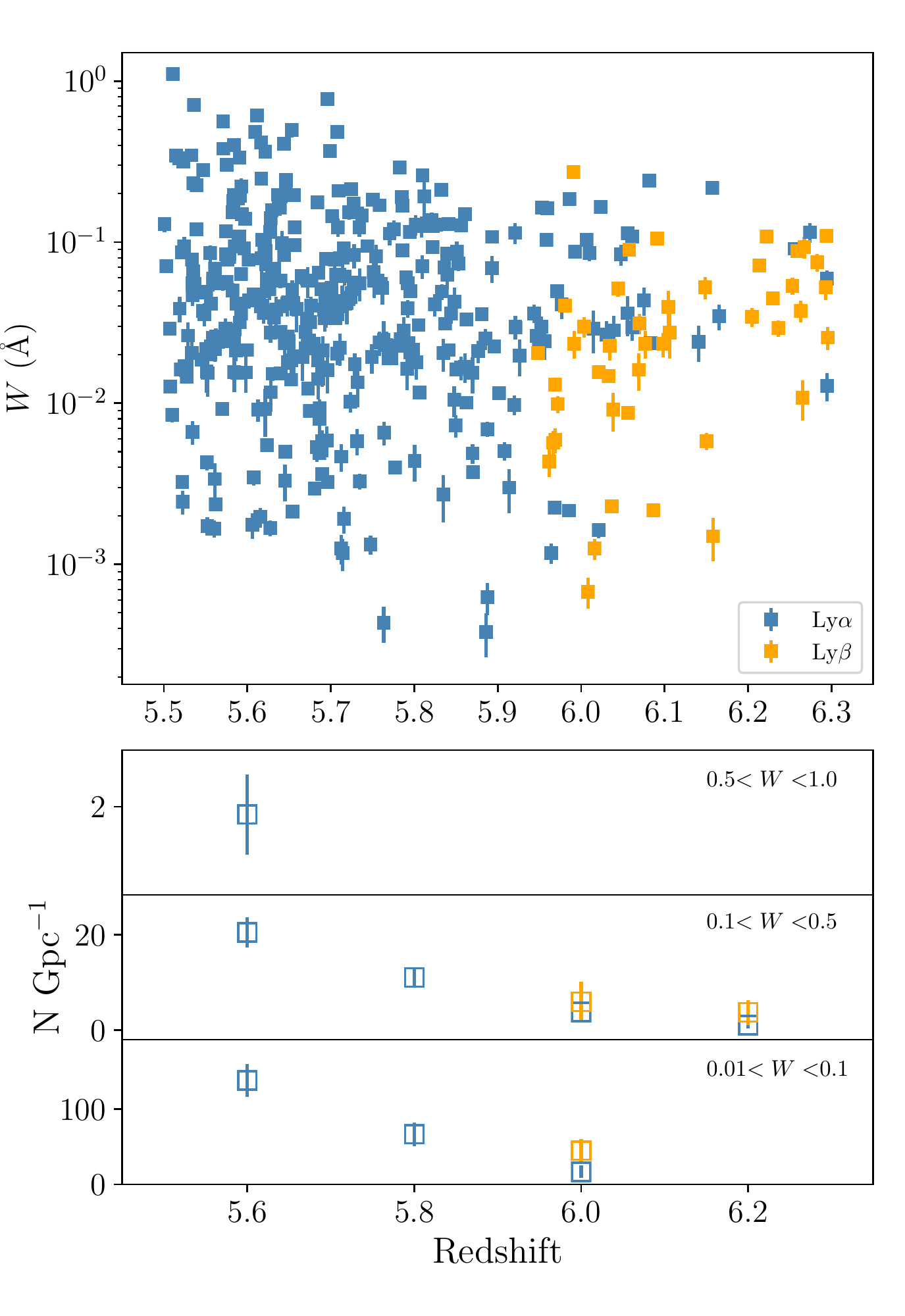} 
\caption{$Top$: The redshift and strength distribution of transmission spikes ($> 3 \sigma$) identified at $z > 5.5$ from all sightlines in our sample. The Ly$\alpha$ (369) and Ly$\beta$ (45) transmission spikes are identified within the Ly$\alpha$ and Ly$\beta$ forests, respectively. The strength $W$ of a transmission spike is measured by integrating transmitted flux over the wavelength range at rest frame. The occurrences of high redshift spikes suggest highly ionized IGM at redshift up to $z \sim 6.3$.
$Bottom$: The number densities along sightline of transmission spike at different $W$ and redshift ($\Delta z = 0.2$) bins.
The obvious decline of number densities towards high redshift is shown at all strength bins, consistent with an increasing neutral fraction.}
\label{fig:spike}
\end{figure}

We first find all peaks at $z_{\rm peak} > 5.5$ above 2$\sigma$ level in the 1D spectra. Then we exclude all one-pixel detections and visually inspect the remaining peaks in the 2D image to confirm the reality of each spike. In total, 389 Ly$\alpha$ and 50 Ly$\beta$ transmission spikes have been identified from all quasar sightlines. We list all these spikes in Table \ref{tab:spike-alpha} and \ref{tab:spike-beta} in Appendix B. The spectra of all $z > 5.9 $ spikes are also shown in Figure \ref{fig:spike-alpha} and \ref{fig:spike-beta} in Appendix B. Limited by the narrow Ly$\beta$ forests, our sample can only cover Ly$\beta$ spikes at $z > 5.9$. 

We define the strength $W$ of a transmission spike using its integrated transmitted flux over the wavelength range of this spike. The wavelength range is in rest-frame. Therefore, the strength is in units of Angstrom. We first use the direct sum of the transmitted flux rather than result from Gaussian fitting, because some spikes have multiple peaks, especially at $z < 5.8$. The multiple peaks could be due to the relatively low spectral resolution or the continuous transmissions at low $\tau$ region. The edges of wavelength range for the integration are defined as the wavelengths where the S/N of transmitted flux drops to 1$\sigma$ at the blue and red sides. Pixels masked due to sky lines are excluded from the integration. An example is shown in Figure \ref{fig:spike-strength}. 
This definition may result in overestimated spike strength of some multi-peak spikes that are not resolved in current spectra (i.e., one spike may be resolved into few spikes with higher resolution spectroscopy). Such spikes are mainly at $z \lesssim 5.8$ where the IGM transmission is higher. 
We only use this method to describe the total transmitted fluxes that we have obtained from our current spectra. The uncertainty of $\sigma_{\rm W}$ is estimated from the transmitted flux uncertainty with proper error propagation. Among all spikes, there are 369 Ly$\alpha$ spikes and 45 Ly$\beta$ spikes with $> 3 \sigma$ detection. We plot their redshift and strength distributions in Figure \ref{fig:spike}.

We calculate the number density along sightline of transmission spikes at different strength $W$ and redshift bins. 
We first estimate the $3 \sigma$ detection limit of transmission spike at each sightline based on its S/N and resolution within the Ly$\alpha$ and Ly$\beta$ forest windows. 
We then select all sightlines that have the detection limits below the lower limit of each $W$ bin and calculate the total path length in comoving Gpc at each redshift bin ($\Delta z =0.2$) based on the selected sightlines.
Since there is only one sightline (J0100+2802) that is deep enough to reach the limit of $W<0.001$, we start our analysis from $W$=0.01 and divide spikes into three $W$ bins as shown in Figure \ref{fig:spike}.
The densities at different strength bins are all declining rapidly with increasing redshift (one order of magnitude from $z\sim5.5$ to $z\sim6.3$), consistent with an increasing neutral fraction over redshift. 
The evolution of spikes over strength has the trend of decreasing density towards high $W$ bin at $z < 6.0$, while at $z>6.0$ the evolution is not clear, which might be caused by the large scatter of IGM optical depth at this redshift range and also the limit of spectral quality. A large dataset of high S/N and high resolution spectra will enable us to better explore the $z>6.0$ transmission spikes evolution in the future.

In addition to the directly integrated strength $W$, we also quantify the strength by Gaussian fitting and compare it with the strength from direct integration described above. We fit each spike with one component Gaussian profile and calculate the equivalent width (EW) by integrating the best-fit (via $\chi^{2}$ minimization) Gaussian profile to describe the spike strength (see Figure \ref{fig:spike-strength}). We have only done the fitting for spikes at $z \ge 5.9$ since more continuous transmissions appear at lower redshift. In addition, the Gaussian fitting procedure does not work well for multi-peak spikes (e.g., the spike at $z=5.99$ in the sightline J0100+2802). When we perform a direct integration of transmitted flux, the pixels masked due to sky lines are not taken into account, while the integration of Gaussian profile includes the masked region. Thus the EW of Gaussian will be higher than the strength from numerical integration if the spike is significantly affected by sky lines (e.g., the spike at $z = 6.02$ in sightline J0305--3150). Most of the EWs are consistent with the strengths from the numerical integration of integrated transmitted flux. All measurements of EW are also listed in Table \ref{tab:spike-alpha} and \ref{tab:spike-beta}.

\begin{figure}
\centering 
\epsscale{1.25}
\plotone{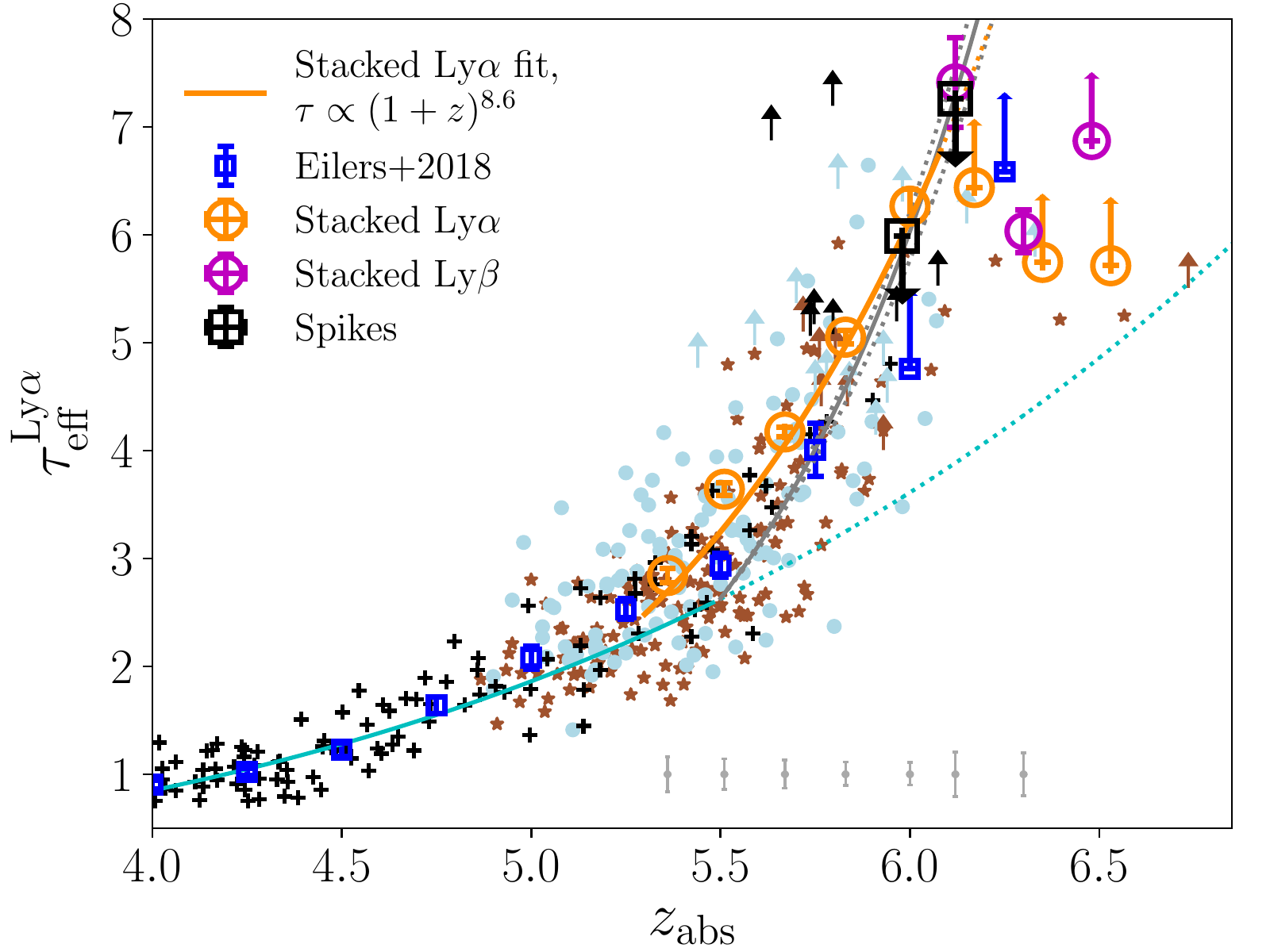} 
\caption{The constrains (black squares) on Ly$\alpha$ effective optical depth estimated using the transmission spikes at $z > 5.9$. Since we are assuming that these transmission spikes could represent all transmitted flux, these two points are treated as upper limits of effective optical depth. All other markers are the same to those in Figure \ref{fig:tau-alpha-beta}. The upper limits from transmission spikes agree with our best-fit evolution model from Ly$\alpha$ measurement.}
\label{fig:tau-spike}
\end{figure}

If we assume that these transmission spikes represent all transmitted flux at $z > 5.9$ in all sightlines (i.e., transmitted flux is equal to zero elsewhere), we can then estimate a lower limit of the mean transmitted flux or an upper limit of the effective optical depth. 
We estimate the upper limits of Ly$\alpha$ effective optical depth using all $> 3 \sigma$ Ly$\alpha$ transmission spikes at $z > 5.9$, with 45 Ly$\alpha$ spikes used in total.
We calculate the mean transmitted flux in two 50 cMpc $h^{-1}$ bins starting at $z = 5.9$ of each spectrum and assume that the identified spikes are the only fluxes in these two bins in all sightlines. Then at each bin we average the mean transmitted fluxes from all sightlines. After that, we obtain the corresponding effective optical depth (black squares in Figure \ref{fig:tau-spike}) at the two bins, $z= 5.98$ and 6.12. 
The upper limits of effective optical depths are slightly above the best-fit evolution model from Ly$\alpha$ measurements as detailed in Section 4.1, which supports the best-fit evolution model derived from our Ly$\alpha$ measurements. On the other hand, the agreement also suggests that our identification of these transmission spikes is highly complete, at least for the strong spikes that dominate the mean transmitted fluxes. 

These strong transmission spikes at high redshift, in particular $6< z <6.3$, could represent the inhomogeneous IGM optical depth. 
The existence of these transmission spikes also suggest the occurrences of highly ionized regions at redshift up to $z\sim$6.3. The evolution in number density of these transmission spikes suggests a rapid transition phase at the end of the reionization.
In addition, the spike properties (e.g., number, height, and width) have been found to correlate with the ionized fraction, gas density, and/or the gas temperature in the ionized regions \citep{garaldi19, gaikwad20}. A large sample of high redshift transmission spikes from a systematic search, comparing with the simulations of IGM, will reveal the process of the reionization in details. 

\section{Neutral Hydrogen Fraction}
Methods to estimate the neutral hydrogen fraction at $z \gtrsim 5$ through high redshift quasar spectra include: (i) using the Ly$\alpha$ or Ly$\beta$ transmissions (i.e., $\tau_{\rm eff}$), which depends on the IGM model \citep{fan06, becker15}; (ii) using the covering fraction of dark pixels, which is model-independent but only provides upper limits \citep{mcgreer11, mcgreer15}; (iii) measuring damping wings features from the highest redshift quasars, which requires quasars residing in a significant neutral environment \citep{mortlock11,greig17,banados18,davies18b}; (iv) measuring the properties of quasar proximity zones \citep{fan06, carilli10, calverley11, venemans15}, which could depend strongly on the quasar lifetime \citep{eilers17,davies20}. These measurements compliment the results obtained by measuring the suppression of Ly$\alpha$ emission in galaxies \citep{kashikawa06,konno18,mason18,mason19,ouchi10}.
Several previous studies have employed these methods and provided different measurements/constraints on the neutral hydrogen fraction up to $z \sim 7.5-8$. 

In this work, using our measurements of $\tau_{\rm eff}^{\rm Ly\alpha}$ and the hydrodynamical simulations of a uniform UVB model in Section 4.2, we are able to directly measure/limit the volume-averaged neutral hydrogen fraction. As described in Section 4.2, we obtain a scaling factor $A_0$ for each redshift bin by matching the median $\tau$ of simulated Ly$\alpha$ forest skewers to the median $\tau$ of our observations. Using the same mimicked skewers, similar scaling factors were also determined for the 16th and 84th percentiles of the $\tau_{\rm eff}^{\rm Ly\alpha}$ at each redshift bin, which are used to estimate the uncertainties of neutral fractions. The corresponding scaling factors for 16th and 84th percentiles $\tau$ are 0.52, 0.76, 0.74, 0.85, 0.77 at $z=5.4, 5.6, 5.8, 6.0, 6.2$, respectively and 0.74, 1.03, 1.05 at three low redshift bins (no 84th scaling factors at $z=6.0$ and 6.2). As discussed in Section 4.2, in the two highest redshift bins, the CDFs and thus the scaling factors are strongly affected by lower limits of $\tau_{\rm eff}^{\rm Ly\alpha}$, therefore we consider the $\langle f_{\rm HI} \rangle$ in these two bins as lower limits here.
The (inverse of the) scaling factors can then be applied to the uniform UVB in the simulations to recover the implied UVB of the observed IGM, similar to \citet{bolton07} and \citet{becker13} although they matched the mean Ly$\alpha$ forest flux instead of the median of the distribution. We then measure $\langle f_{\rm HI}\rangle$ directly from the distribution of neutral hydrogen in the simulation skewers after applying the rescaled UVBs.

\begin{figure}
\centering 
\epsscale{1.1}
\plotone{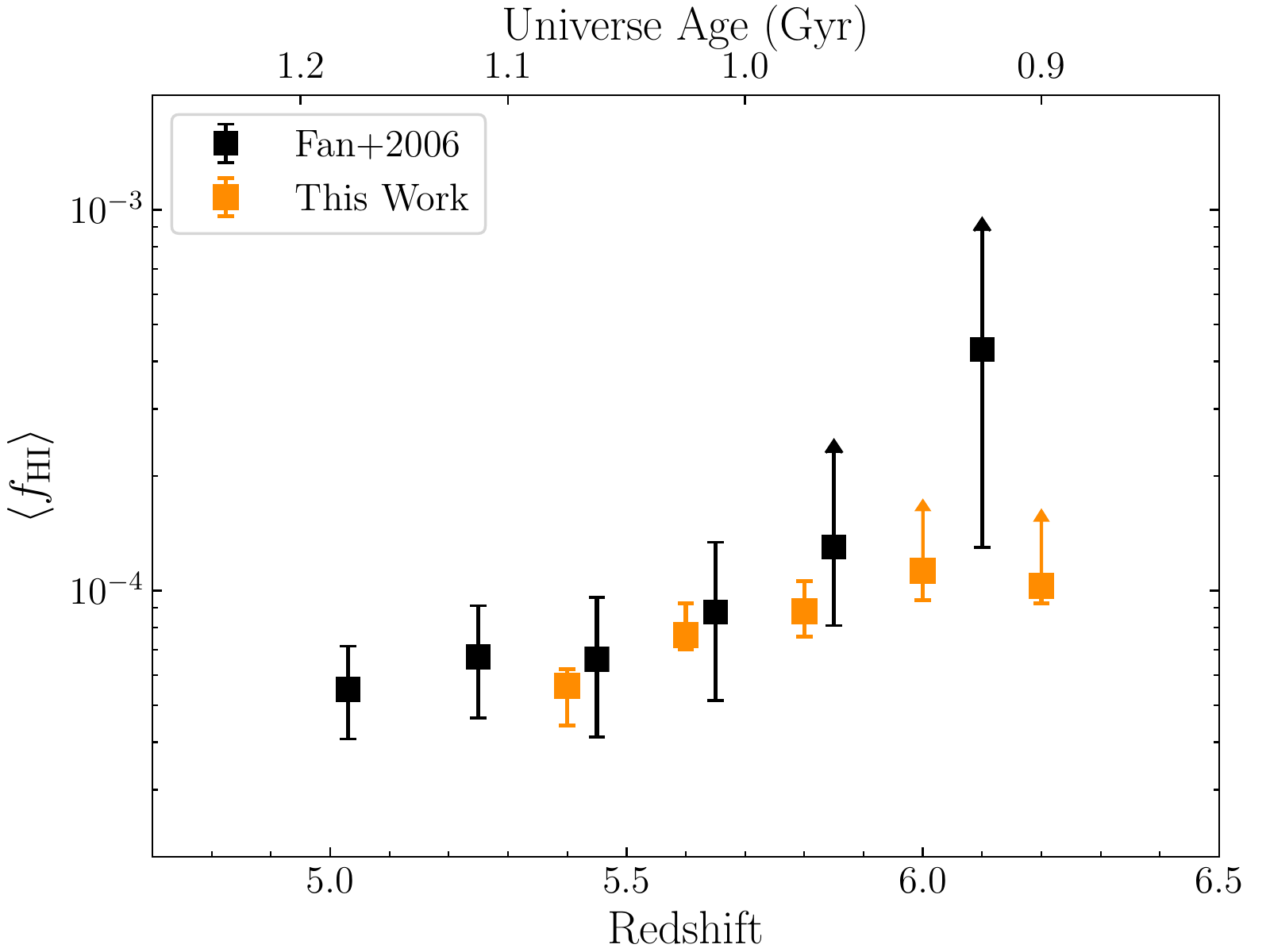} 
\caption{The neutral hydrogen fraction generated from our measurements of $\tau_{\rm eff}^{\rm Ly\alpha}$ and the uniform UVB model described in Section 4.2, compared with previous measurements/limits from \cite{fan06}. Our estimates place new constraints on the neutral fraction at $z < 6$. These estimates are consistent with \cite{fan06} within the uncertainties. The errors on neutral fractions are estimated based on the scaling factors from 16th and 84th percentile of observed $\tau_{\rm eff}^{\rm Ly\alpha}$ at each bins, representing the scatters of $\tau_{\rm eff}^{\rm Ly\alpha}$. We should note that the neutral fraction measurements/constraints estimated based on the measurements of effective optical depth and the uniform UVB simulation model are ambiguous and our results do not rule out a $\langle f_{\rm HI} \rangle$ $\sim$ 10-20\% at $z \sim$ 6.}
\label{fig:xhi}
\end{figure}

Our neutral fraction measurements are shown in Figure \ref{fig:xhi}, compared with the measurements/constraints from \cite{fan06}, who calculated $\langle f_{\rm HI} \rangle$ from their effective optical depth measurements and a parametric form for the IGM density distribution. Our new constraints are consistent with the results from \cite{fan06} within uncertainties at $z\lesssim6$, with an increase in the neutral fraction by a factor of two between $z\sim5.5$ and $z\sim6$. At $z \sim 6$ the $\langle f_{\rm HI} \rangle$ should be $\gtrsim 10^{-4}$. At $z>6$, our measurements of $\tau_{\rm eff}^{\rm Ly\alpha}$ start to be affected/dominated by the lower limits and thus could only provide loose constraints. The high limit at $z=6.1$ in \citep{fan06} was estimated based on the optical depths only from three sightlines among which one sightline includes a complete Gunn-Peterson trough and has negative observed transmitted flux. Consequently, it could be biased to high $\tau$. Our apparently less stringent limits at $z=6$ and $z=6.2$ should be more representative of the IGM as a whole due to our much larger sample sizes. To improve the measurements/constraints at $z > 6$, higher quality spectra and larger samples are required. 

We stress that our neutral fraction constraints do not rule out an incomplete reionization epoch below $z\sim6$ \citep{kulkarni19,keating20,choudhury20}, as we have computed them assuming a fully ionized IGM with a uniform UVB. However, as long as the true $\langle f_{\rm HI} \rangle$ is still relatively small, they represent a good approximation to the neutral fraction inside of the majority of the volume of the IGM which is highly ionized (cf. \citealt{mcgreer15}). We leave constraints on fluctuating scenarios to future work.

\section{Summary}
In this paper, we report the measurements of Ly$\alpha$ optical depth by computing the effective optical depth of 32 quasars in the redshift range of $6.3 < z <= 7.0$. The new sample of quasar sightlines is constructed based on $z > 6.3$ quasars from wide field high redshift quasar surveys during the past few years \citep{fan01b,fan03, mortlock11, venemans15, wu15, mazzucchelli17, wang19, yang19}. Among 22 $z > 6.5$ quasars, 13 are new quasars discovered from our new $z \sim 7$ quasar survey \citep{wang19,yang19}. This new sample is the largest collection of $z > 6.3$ quasar sightline. We construct a data set of deep optical spectra of these 32 quasars with spectra obtained from VLT/X-Shooter, Keck/DEIMOS, Keck/LRIS, LBT/MODS, Gemini/GMOS, and MMT/BINOSPEC. We measure/constrain Ly$\alpha$ effective optical depth and neutral hydrogen fraction using these sightlines and compare our results with previous works and simulated IGM models. Our main results are summarized as follows. 

\begin{itemize}
\item We measure the Ly$\alpha$ effective optical depth in Ly$\alpha$ forests of individual sightlines at fixed 50 cMpc $h^{-1}$ bins, covering the redshift range of $5.25 < z < 6.7$. Our work expands the dataset of $\tau_{\rm eff}^{\rm Ly\alpha}$ measurements towards higher redshift ($z > 6$) than previous works. These individual measurements are following the trend of the optical evolution from previous studies and also show the increasing scatter at $5.5 \le z \le 6$, supporting a spatially inhomogeneous reionization process.

\item We stack 1D transmitted flux in Ly$\alpha$ forests of all sightlines and measure the effective optical depth from the stacked spectrum. We obtain measurements of $\tau_{\rm eff}^{\rm Ly\alpha}$ up to $z=6.0$. The combination of our new measurements and two mean optical depth data points from \cite{eilers18} at $5.3 < z \le 6.0$ yields a best-fit of the Ly$\alpha$ effective optical depth evolution, $\tau \propto (1+z)^{8.6\pm 1.0}$. 

\item We compute the CDFs of $\tau_{\rm eff}^{\rm Ly\alpha}$ at five bins at $5.3 < z < 6.3$. Our CDFs are marginally consistent with the previous observations. The discrepancy between our measurements and the CDFs from uniform UVB model simulations support the disagreement between observations and uniform UVB model, suggesting the requirement of fluctuations in radiation background, temperature, or a combination of them. 

\item We measure the effective optical depth in Ly$\beta$ forests of each sightline at fixed 30 cMpc $h^{-1}$ bins, including both observed and pure $\tau_{\rm eff}^{\rm Ly\beta}$. The observed effective optical depths, $\tau_{\rm eff}^{\rm Ly\beta, obs}$, show a trend of rapid increasing over redshift but the exact evolution needs to be constrained with more sightlines. The larger scatter when compared to the prediction of the uniform UVB model further supports a fluctuating UVB or/and fluctuating temperature models. We convert the pure $\tau_{\rm eff}^{\rm Ly\beta}$ to the Ly$\alpha$ effective optical depth using a conversion factor. The Ly$\beta$-based measurements of $\tau_{\rm eff}^{\rm Ly\alpha}$ are in agreement with the results from Ly$\alpha$ forests, although this method is highly dependent on the fluctuations of the foreground Ly$\alpha$ absorptions and the conversion factor between $\tau_{\rm eff}^{\rm Ly\alpha}$ and $\tau_{\rm eff}^{\rm Ly\beta}$.

\item We use our new sample to search for high redshift transmission spikes in both Ly$\alpha$ and Ly$\beta$ forests. 389 Ly$\alpha$ transmission spikes and 50 Ly$\beta$ transmission spikes are identified up to redshift $z = 6.29$. We estimate the upper limits of Ly$\alpha$ effective optical depth at $z = 5.98$ and 6.12 using 45 Ly$\alpha$ transmission spikes. The upper limits are sightly higher than our best-fit evolution model from $\tau_{\rm eff}^{\rm Ly\alpha}$, providing additional tight constraints on the Ly$\alpha$ optical depth and suggesting a high completeness of our transmission spike sample. The existence of these transmission spikes suggest the occurrences of highly ionized regions at redshift up to $z\sim6.3$. The evolution in number density of these high-redshift transmission spikes suggests a rapid transition phase at the end of the reionization.

\item The neutral hydrogen fraction estimated from Ly$\alpha$ effective optical depth measurements and the hydrodynamical simulations assuming uniform UVB model is consistent with the results from \cite{fan06} at $z < 6$. At $z \sim 6$, we obtain a $\langle f_{\rm HI} \rangle$ $\gtrsim 10^{-4}$.

\end{itemize}

Further studies by combing this optical dataset with NIR spectral data from our on-going NIR spectroscopic survey will allow accurate measurements of quasar proximity zone size and improved intrinsic continuum fitting. The cutoff of 1176 \AA\ used in this work is conservative and thus the estimates of proximity zone sizes will extend the available Ly$\alpha$/Ly$\beta$ forest windows towards higher redshift, which will enable us to explore the optical depth at the high-redshift end with more measurements and search for transmission spikes at $z > 6.3$. The study of proximity zone will also constrain the quasar life time and the IGM evolution.
We are continue to expand the sample with observations of newly discovered quasars and higher quality optical spectra, which will provide significant extension of this work and allow the investigation of IGM neutral fraction using dark pixels.


\acknowledgments
J. Yang, X. Fan and M. Yue acknowledge the supports from the US NSF grant AST 15-15115, AST 19-08284 and NASA ADAP Grant NNX17AF28G.
F. Wang acknowledges support by NASA through the NASA Hubble Fellowship grant \#HST-HF2-51448.001-A awarded by the Space Telescope Science Institute, which is operated by the Association of Universities for Research in Astronomy, Incorporated, under NASA contract NAS5-26555.
E.P. Farina acknowledge support from the ERC Advanced Grant 740246 (Cosmic Gas).
X.-B. Wu thanks the supports by the National Key R\&D Program of China (2016YFA0400703) and the National Science Foundation of China (11533001, 11721303).
F. Pacucci acknowledges support from the Yale Keck program No. Y144 and from the Black Hole Initiative at Harvard University, which is funded by grants from the John Templeton Foundation and the Gordon and Betty Moore Foundation. 

Some of the data presented in this paper were obtained at the W.M. Keck Observatory, which is operated as a scientific partnership among the California Institute of Technology, the University of California and the National Aeronautics and Space Administration.
The Observatory was made possible by the generous financial support of the W. M. Keck Foundation.
The authors wish to recognize and acknowledge the very significant cultural role and reverence that the summit of Maunakea has always had within the indigenous Hawaiian community. We are most fortunate to have the opportunity to conduct observations from this mountain.
This work is based in part on observations made with ESO telescopes at the La Silla Paranal Observatory under program IDs 084.A-0360(A), 086.A-0162(A), 087.A-0607(A), 087.A-0890(A), 088.A-0897(A), 096.A-0095(A), 096.A-0418(A), 097.B-1070(A), 098.A-0444(A), 098.A-0527(A), 098.B-0537(A), 0100.A-0446(A), 0100.A-0625(A), 0102.A-0154(A), 0103.A-0423(A).
This paper also uses data based on observations obtained at the Gemini Observatory, which is operated by the Association of Universities for Research in Astronomy, Inc., under a cooperative agreement with the NSF on behalf of the Gemini partnership: the National Science Foundation (United States), the National Research Council (Canada), CONICYT (Chile), Ministerio de Ciencia, Tecnolog\'{i}a e Innovaci\'{o}n Productiva (Argentina), and Minist\'{e}rio da Ci\^{e}ncia, Tecnologia e Inova\c{c}\~{a}o (Brazil).
We acknowledge the use of the MMT, LBT,  and VLT telescopes. We acknowledge the use of the PypeIt data reduction package.

%

\vspace{5mm}
\facilities{Gemini(GMOS), Keck(DEIMOS,LRIS), LBT(MODS), MMT(BINOSPEC),VLT,(X-Shooter)}


\software{PypeIt (\url{https://zenodo.org/record/3743493})}



\appendix
\section{Effective Optical Depth Measurements within Ly$\alpha$ and Ly$\beta$ Forests}
We plot all measurements of the mean transmitted flux in the Ly$\alpha$ forest with the bin size of 50 comoving Mpc $h^{-1}$ in Figure \ref{fig:forest1} and the Ly$\beta$ forest with the bin size of 30 comoving Mpc $h^{-1}$ in Figure \ref{fig:betaforest}. The mean fluxes measured at each bin are listed in Table \ref{tab:F-alpha} and \ref{tab:F-beta}.
The mean flux within the Ly$\beta$ forest is the observed transmitted flux including the absorption from foreground Ly$\alpha$.

\startlongtable
\begin{deluxetable}{l l l c l}
\tablecaption{Measurements of mean transmitted flux in the Ly$\alpha$ forests.}
\tabletypesize{\footnotesize}
\tablewidth{\textwidth}
\tablehead{
\colhead{Object} &
\colhead{$z_{\rm em}$} &
\colhead{$z_{\rm abs}$} &
\colhead{$\langle F\rangle$} &
\colhead{$\sigma_{\langle F\rangle}$} 
}
\startdata
J0252$-$0503  &  7.0  &  6.09  &  $-$0.0026  &  0.003 \\
 &  &  6.27  &  $-$0.0059  &  0.0029 \\
 &  &  6.45  &  $-$0.0068  &  0.0023 \\
 &  &  6.64  &  0.0016  &  0.0039 \\
J0020$-$3653  &  6.834  &  5.79  &  0.0088  &  0.003 \\
 &  &  5.95  &  0.0171  &  0.0031 \\
 &  &  6.12  &  0.0006  &  0.0038 \\
 &  &  6.30  &  $-$0.0025  &  0.0037 \\
 &  &  6.48  &  0.002  &  0.0028 \\
J0319$-$1008  &  6.83  &  5.78  &  0.027  &  0.0048 \\
 &  &  5.95  &  0.0043  &  0.006 \\
 &  &  6.12  &  $-$0.0014  &  0.0037 \\
 &  &  6.30  &  $-$0.0033  &  0.0079 \\
 &  &  6.48  &  $-$0.0011  &  0.0037 \\
J0411$-$0907  &  6.81  &  5.77  &  0.0256  &  0.0039 \\
 &  &  5.93  &  0.0107  &  0.0045 \\
 &  &  6.10  &  0.0027  &  0.0036 \\
 &  &  6.28  &  0.0104  &  0.0058 \\
 &  &  6.46  &  0.0059  &  0.0045 \\
J0109$-$3047  &  6.7909  &  5.75  &  0.0321  &  0.0056 \\
 &  &  5.92  &  $-$0.0206  &  0.0061 \\
 &  &  6.08  &  0.0075  &  0.0056 \\
 &  &  6.26  &  $-$0.016  &  0.0069 \\
 &  &  6.44  &  $-$0.0236  &  0.0061 \\
J0218+0007  &  6.77  &  5.73  &  0.04  &  0.0065 \\
 &  &  5.90  &  $-$0.0108  &  0.0099 \\
 &  &  6.07  &  0.0103  &  0.0068 \\
 &  &  6.24  &  $-$0.0022  &  0.0101 \\
 &  &  6.42  &  0.0063  &  0.0075 \\
J1104+2134  &  6.74  &  5.71  &  0.0354  &  0.0031 \\
 &  &  5.87  &  0.0252  &  0.0027 \\
 &  &  6.04  &  0.0197  &  0.002 \\
 &  &  6.21  &  0.0238  &  0.0022 \\
 &  &  6.40  &  0.0111  &  0.0018 \\
J0910+1656  &  6.72  &  5.69  &  0.0059  &  0.0038 \\
 &  &  5.85  &  0.0085  &  0.0055 \\
 &  &  6.02  &  $-$0.0011  &  0.0044 \\
 &  &  6.20  &  $-$0.0007  &  0.0056 \\
 &  &  6.38  &  $-$0.0061  &  0.0055 \\
J0837+4929  &  6.71  &  5.68  &  0.0184  &  0.0037 \\
 &  &  5.85  &  0.0034  &  0.0052 \\
 &  &  6.01  &  0.0058  &  0.0046 \\
 &  &  6.19  &  0.0112  &  0.006 \\
 &  &  6.37  &  0.0046  &  0.0063 \\
J1048$-$0109  &  6.6759  &  5.66  &  0.0429  &  0.0071 \\
 &  &  5.82  &  0.0017  &  0.0065 \\
 &  &  5.98  &  $-$0.0119  &  0.0065 \\
 &  &  6.16  &  $-$0.0278  &  0.0094 \\
 &  &  6.33  &  $-$0.008  &  0.0081 \\
J2002$-$3013  &  6.67  &  5.65  &  0.0348  &  0.0035 \\
 &  &  5.81  &  0.0207  &  0.0031 \\
 &  &  5.98  &  0.0078  &  0.0028 \\
 &  &  6.15  &  0.0058  &  0.0028 \\
 &  &  6.33  &  0.0088  &  0.0036 \\
J2232+2930  &  6.658  &  5.64  &  0.0103  &  0.0025 \\
 &  &  5.80  &  0.0015  &  0.0027 \\
 &  &  5.97  &  0.0008  &  0.0025 \\
 &  &  6.14  &  $-$0.0038  &  0.0029 \\
 &  &  6.32  &  $-$0.0051  &  0.0028 \\
J1216+4519  &  6.654  &  5.64  &  0.0378  &  0.0043 \\
 &  &  5.80  &  0.0252  &  0.0038 \\
 &  &  5.96  &  $-$0.0041  &  0.0035 \\
 &  &  6.14  &  0.0074  &  0.0037 \\
 &  &  6.31  &  0.0003  &  0.0053 \\
J2102$-$1458  &  6.648  &  5.63  &  0.0021  &  0.0045 \\
 &  &  5.79  &  $-$0.0046  &  0.005 \\
 &  &  5.96  &  0.0033  &  0.005 \\
 &  &  6.13  &  $-$0.008  &  0.0051 \\
 &  &  6.31  &  0.0064  &  0.0053 \\
J0024+3913  &  6.621  &  5.61  &  0.0226  &  0.0027 \\
 &  &  5.77  &  0.0332  &  0.003 \\
 &  &  5.93  &  0.0091  &  0.0028 \\
 &  &  6.11  &  0.0213  &  0.0032 \\
 &  &  6.28  &  0.0178  &  0.0029 \\
J0305$-$3150  &  6.6145  &  5.61  &  0.0162  &  0.0042 \\
 &  &  5.76  &  $-$0.0005  &  0.0035 \\
 &  &  5.93  &  $-$0.0003  &  0.004 \\
 &  &  6.10  &  0.0  &  0.0047 \\
 &  &  6.28  &  $-$0.0132  &  0.005 \\
J1526$-$2050  &  6.5864  &  5.58  &  0.0796  &  0.0064 \\
 &  &  5.74  &  0.0346  &  0.0027 \\
 &  &  5.90  &  0.0377  &  0.0028 \\
 &  &  6.07  &  $-$0.0012  &  0.0019 \\
 &  &  6.25  &  0.0009  &  0.0017 \\
J2132+1217  &  6.585  &  5.58  &  0.0227  &  0.0022 \\
 &  &  5.74  &  0.0012  &  0.0011 \\
 &  &  5.90  &  0.0068  &  0.0016 \\
 &  &  6.07  &  0.0016  &  0.0014 \\
 &  &  6.25  &  0.0035  &  0.0016 \\
J1135+5011  &  6.58  &  5.58  &  0.0514  &  0.0068 \\
 &  &  5.74  &  0.0544  &  0.0056 \\
 &  &  5.90  &  0.033  &  0.0059 \\
 &  &  6.07  &  0.0172  &  0.0056 \\
 &  &  6.24  &  0.0013  &  0.005 \\
J0226+0302  &  6.5412  &  5.55  &  0.0959  &  0.0075 \\
 &  &  5.70  &  0.0195  &  0.0016 \\
 &  &  5.86  &  0.0061  &  0.0014 \\
 &  &  6.03  &  $-$0.0008  &  0.0011 \\
 &  &  6.21  &  $-$0.0028  &  0.0012 \\
J0148$-$2826  &  6.54  &  5.55  &  0.0272  &  0.0104 \\
 &  &  5.70  &  0.0074  &  0.0067 \\
 &  &  5.86  &  0.0302  &  0.0111 \\
 &  &  6.03  &  $-$0.0039  &  0.0084 \\
 &  &  6.21  &  0.0128  &  0.0109 \\
J0224$-$4711  &  6.526  &  5.53  &  0.04  &  0.0043 \\
 &  &  5.69  &  0.0052  &  0.002 \\
 &  &  5.85  &  0.0032  &  0.0028 \\
 &  &  6.02  &  $-$0.003  &  0.0026 \\
 &  &  6.19  &  $-$0.0031  &  0.0039 \\
J1629+2407  &  6.476  &  5.49  &  0.0434  &  0.0041 \\
 &  &  5.65  &  0.0375  &  0.0031 \\
 &  &  5.81  &  0.0072  &  0.0024 \\
 &  &  5.97  &  0.007  &  0.0024 \\
 &  &  6.14  &  0.0179  &  0.003 \\
J2318$-$3113  &  6.4435  &  5.47  &  0.0596  &  0.0054 \\
 &  &  5.62  &  0.0031  &  0.0026 \\
 &  &  5.78  &  0.0099  &  0.0024 \\
 &  &  5.94  &  $-$0.007  &  0.0029 \\
 &  &  6.11  &  $-$0.0049  &  0.0027 \\
J0045+0901  &  6.42  &  5.45  &  0.0585  &  0.0055 \\
 &  &  5.60  &  0.0274  &  0.0033 \\
 &  &  5.76  &  0.0164  &  0.0028 \\
 &  &  5.92  &  0.0031  &  0.0032 \\
 &  &  6.09  &  0.0167  &  0.0036 \\
J1148+5251  &  6.4189  &  5.43  &  0.0255  &  0.002 \\
 &  &  5.58  &  0.0185  &  0.0013 \\
 &  &  5.74  &  0.0044  &  0.0003 \\
 &  &  5.90  &  0.0021  &  0.0003 \\
J1036$-$0232  &  6.3809  &  5.41  &  0.0428  &  0.0038 \\
 &  &  5.57  &  0.0228  &  0.0023 \\
 &  &  5.72  &  0.0033  &  0.0012 \\
 &  &  5.89  &  0.0083  &  0.0019 \\
 &  &  6.05  &  0.0008  &  0.0017 \\
J1152+0055  &  6.3643  &  5.40  &  $-$0.007  &  0.0069 \\
 &  &  5.55  &  $-$0.0049  &  0.0069 \\
 &  &  5.71  &  $-$0.0145  &  0.0052 \\
 &  &  5.87  &  $-$0.0044  &  0.0072 \\
 &  &  6.04  &  0.0016  &  0.0079 \\
J1148+0702  &  6.344  &  5.38  &  0.1219  &  0.0096 \\
 &  &  5.53  &  0.025  &  0.0028 \\
 &  &  5.69  &  0.0276  &  0.0023 \\
 &  &  5.85  &  0.0183  &  0.0023 \\
 &  &  6.02  &  $-$0.0009  &  0.0019 \\
J0142$-$3327  &  6.3379  &  5.38  &  0.0535  &  0.0046 \\
 &  &  5.53  &  0.0378  &  0.0033 \\
 &  &  5.68  &  0.0096  &  0.0017 \\
 &  &  5.85  &  0.0173  &  0.0022 \\
 &  &  6.01  &  $-$0.0018  &  0.0021 \\
J0100+2802  &  6.327  &  5.37  &  0.1309  &  0.01 \\
 &  &  5.52  &  0.0586  &  0.0039 \\
 &  &  5.68  &  0.0143  &  0.0008 \\
 &  &  5.84  &  0.0024  &  0.0002 \\
 &  &  6.00  &  0.0032  &  0.0002 \\
J1030+0524  &  6.308  &  5.35  &  0.1085  &  0.0084 \\
 &  &  5.50  &  0.0616  &  0.0042 \\
 &  &  5.66  &  0.033  &  0.002 \\
 &  &  5.82  &  0.0188  &  0.0011 \\
 &  &  5.98  &  0.0011  &  0.0004 \\
\enddata
\label{tab:F-alpha}
\end{deluxetable}
 
 \startlongtable
\begin{deluxetable}{l l l c l}
\tablecaption{Measurements of mean transmitted flux in the Ly$\beta$ forests.}
\tabletypesize{\footnotesize}
\tablewidth{\textwidth}
\tablehead{
\colhead{Object} &
\colhead{$z_{\rm em}$} &
\colhead{$z_{\rm abs}$} &
\colhead{$\langle F\rangle$} &
\colhead{$\sigma_{\langle F\rangle}$} 
}
\startdata
J0252$-$0503  &  7.0  &  6.68  &  0.0028  &  0.0035 \\
J0020$-$3653  &  6.834  &  6.52  &  0.0025  &  0.0036 \\
J0319$-$1008  &  6.83  &  6.52  &  $-$0.0045  &  0.0045 \\
J0411$-$0907  &  6.81  &  6.50  &  0.0118  &  0.005 \\
J0109$-$3047  &  6.7909  &  6.48  &  $-$0.0435  &  0.0095 \\
J0218+0007  &  6.77  &  6.46  &  0.0228  &  0.0125 \\
J1104+2134  &  6.74  &  6.43  &  0.0118  &  0.0029 \\
J0910+1656  &  6.72  &  6.41  &  0.0079  &  0.0072 \\
J0837+4929  &  6.71  &  6.40  &  0.0122  &  0.0056 \\
J1048$-$0109  &  6.6759  &  6.37  &  $-$0.0256  &  0.0095 \\
J2002$-$3013  &  6.67  &  6.36  &  0.0067  &  0.003 \\
J2232+2930  &  6.658  &  6.35  &  $-$0.0052  &  0.0027 \\
J1216+4519  &  6.654  &  6.35  &  0.0  &  0.0034 \\
J2102$-$1458  &  6.648  &  6.34  &  $-$0.0067  &  0.0053 \\
J0024+3913  &  6.621  &  6.32  &  0.0184  &  0.0024 \\
J0305$-$3150  &  6.6145  &  6.31  &  0.0011  &  0.0045 \\
J1526$-$2050  &  6.5864  &  6.28  &  0.0061  &  0.0017 \\
J2132+1217  &  6.585  &  6.28  &  0.02  &  0.0014 \\
J1135+5011  &  6.58  &  6.28  &  0.0103  &  0.0059 \\
J0226+0302  &  6.5412  &  6.24  &  0.0085  &  0.0014 \\
J0148$-$2826  &  6.54  &  6.24  &  $-$0.015  &  0.0092 \\
J0224$-$4711  &  6.526  &  6.23  &  0.0178  &  0.0028 \\
J1629+2407  &  6.476  &  6.18  &  0.0065  &  0.0026 \\
J2318$-$3113  &  6.4435  &  6.15  &  0.0003  &  0.0035 \\
J0045+0901  &  6.42  &  6.13  &  0.0061  &  0.0039 \\
J1036$-$0232  &  6.3809  &  6.09  &  0.0137  &  0.0022 \\
J1152+0055  &  6.3643  &  6.07  &  0.0216  &  0.0083 \\
J1148+0702  &  6.344  &  6.05  &  0.0127  &  0.0026 \\
J0142$-$3327  &  6.3379  &  6.05  &  0.0082  &  0.0024 \\
J0100+2802  &  6.327  &  6.04  &  0.0195  &  0.0003 \\
J1030+0524  &  6.308  &  6.02  &  0.005  &  0.0006 \\
\enddata
\label{tab:F-beta}
\end{deluxetable}

\begin{figure*}
\centering 
\includegraphics[width=1.0\textwidth]{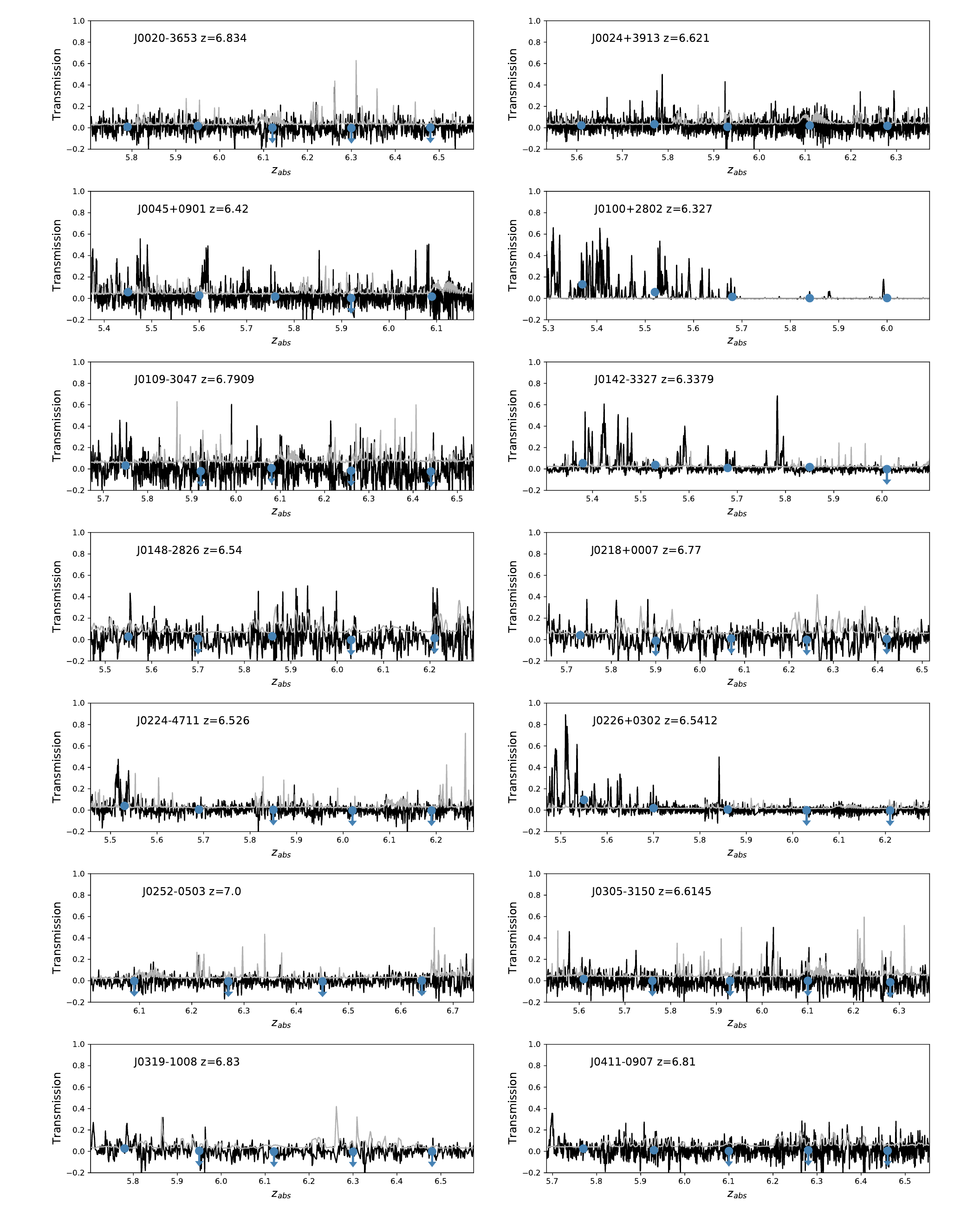}
\caption{Ly$\alpha$ forests and mean transmitted fluxes measured at each bin. The blue points show the measurements/upper limits of $\langle F^{\rm obs}\rangle$ at each bin. The X-Shooter spectra have been binned with 3 pixels here.}
\label{fig:forest1}
\end{figure*}

\addtocounter{figure}{-1}
\begin{figure*}
\centering 
\includegraphics[width=1.0\textwidth]{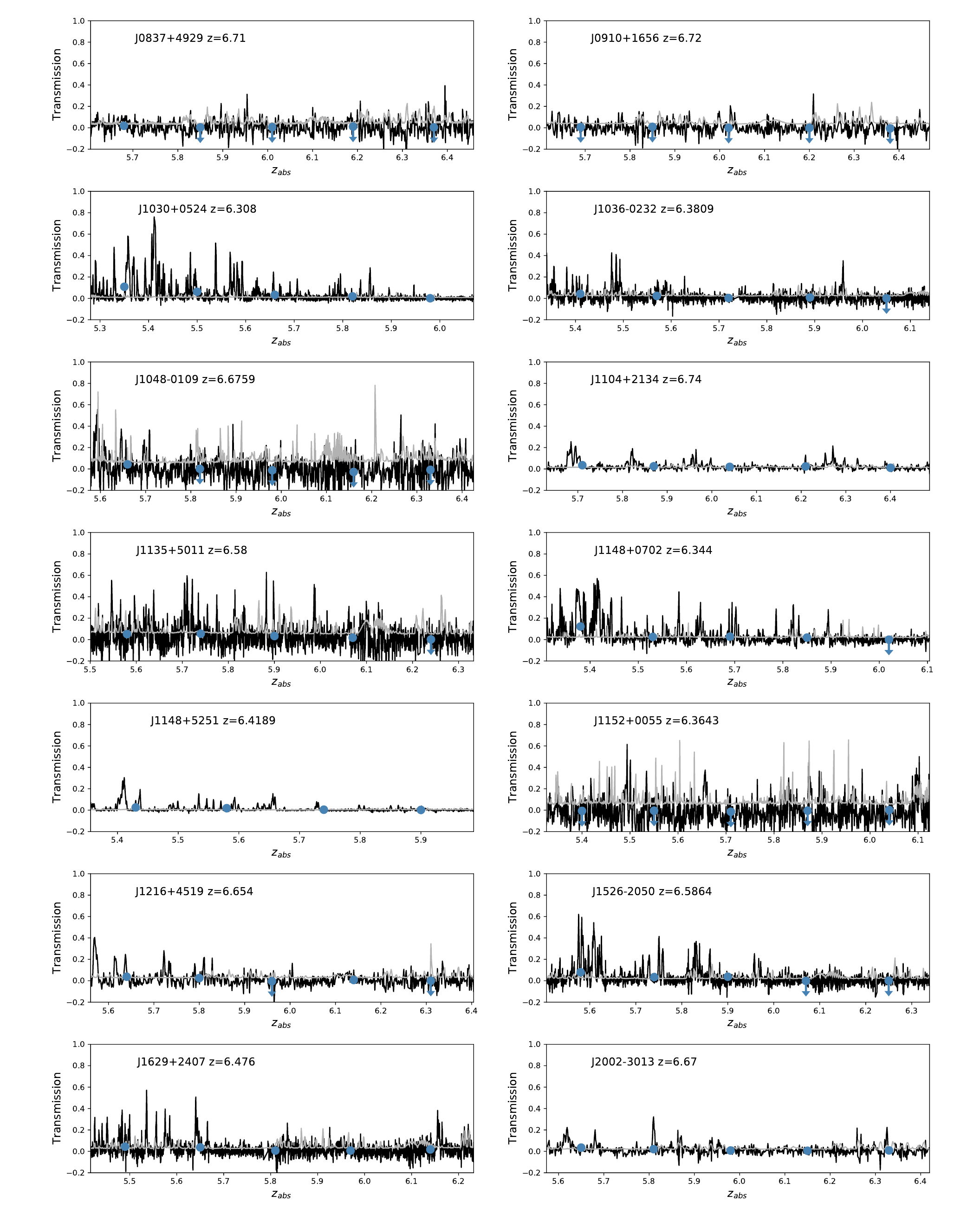}
\caption{Continued.}
\label{fig:forest2}
\end{figure*}

\addtocounter{figure}{-1}
\begin{figure*}
\centering 
\includegraphics[width=1.0\textwidth]{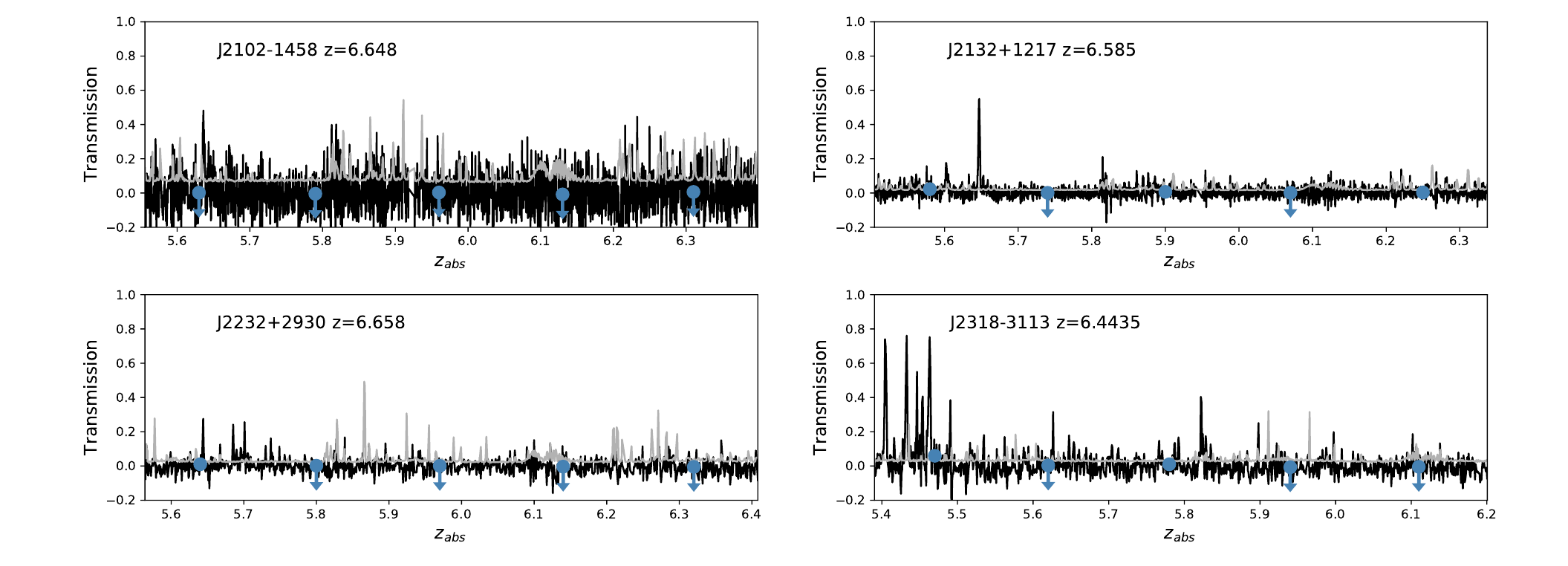}
\caption{Continued.}
\label{fig:forest3}
\end{figure*}

\begin{figure*}
\centering 
\includegraphics[width=1.0\textwidth]{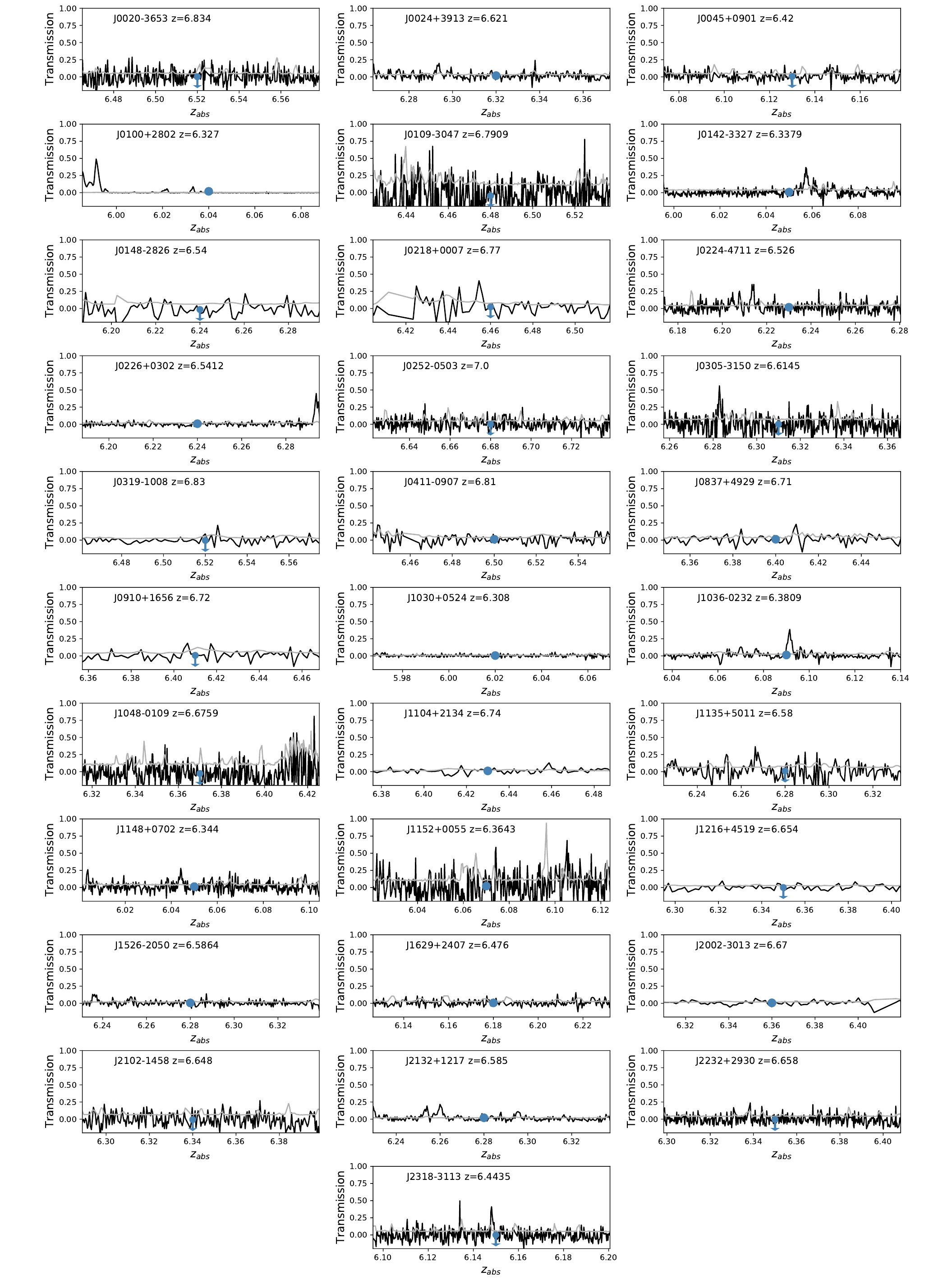}
\caption{Ly$\beta$ forests and mean transmitted fluxes measured at each bin. The mean transmitted flux here is the observed flux $F_{\rm \beta, obs}$.}
\label{fig:betaforest}
\end{figure*}

\clearpage
\section{Catalogs of Transmission Spikes Identified From Our Quasar Sample}
Here we list all transmission spikes identified from all 32 quasars within the Ly$\alpha$ (rest frame 1040 -- 1176 \AA) and Ly$\beta$ (rest frame 975 -- 1176/$\lambda_{\rm \alpha}$ $\times$ $\lambda_{\rm \beta}$ \AA) forests.

\startlongtable
\begin{deluxetable}{l l c l l l l}
\tablecaption{Ly$\alpha$ Transmission Spikes at $z > 5.5$.}
\tabletypesize{\footnotesize}
\tablewidth{\textwidth}
\tablehead{
\colhead{Object} &
\colhead{$z_{\rm em}$} &
\colhead{$z_{\rm spike}$} &
\colhead{$W$} &
\colhead{$\sigma_{W}$} &
\colhead{EW} &
\colhead{$\sigma_{\rm EW}$} 
}
\startdata
J0319--1008  &  6.83  &  5.71  &  0.21  &  0.02 & --- & --- \\
  &   &  5.77  &  0.11  &  0.02 & --- & --- \\
  &   &  5.78  &  0.19  &  0.02 & --- & --- \\
  &   &  5.80  &  0.13  &  0.02 & --- & --- \\
J0411--0907  &  6.81  &  5.70  &  0.37  &  0.02 & --- & --- \\
  &   &  5.71  &  0.08  &  0.01 & --- & --- \\
  &   &  5.73  &  0.08  &  0.01 & --- & --- \\
  &   &  5.85  &  0.09  &  0.01 & --- & --- \\
J0109--3047  &  6.7909  &  5.72  &  0.03  &  0.01 & --- & --- \\
  &   &  5.73  &  0.05  &  0.01 & --- & --- \\
  &   &  5.74  &  0.15  &  0.02 & --- & --- \\
  &   &  5.75  &  0.06  &  0.01 & --- & --- \\
  &   &  5.76  &  0.02  &  0.01 & --- & --- \\
  &   &  5.76  &  0.05  &  0.01 & --- & --- \\
  &   &  5.76  &  0.02  &  0.01 & --- & --- \\
  &   &  6.05  &  0.08  &  0.01  &  0.1  &  0.02 \\
  &   &  6.06  &  0.04  &  0.01  &  0.04  &  0.01 \\
J0218+0007  &  6.77  &  5.65  &  0.50  &  0.05 & --- & --- \\
  &   &  5.81  &  0.19  &  0.06 & --- & --- \\
J1104+2134  &  6.74  &  5.70  &  0.77  &  0.02 & --- & --- \\
  &   &  5.81  &  0.07  &  0.01 & --- & --- \\
  &   &  5.82  &  0.14  &  0.02 & --- & --- \\
  &   &  5.84  &  0.06  &  0.01 & --- & --- \\
  &   &  5.95  &  0.16  &  0.01  &  0.14  &  0.02 \\
  &   &  5.98  &  0.04  &  0.01  &  0.05  &  0.01 \\
  &   &  6.21  &  0.02  &  0.01  &  0.05  &  0.02 \\
  &   &  6.26  &  0.09  &  0.01  &  0.1  &  0.01 \\
  &   &  6.27  &  0.12  &  0.02  &  0.19  &  0.02 \\
J0910+1656  &  6.72  &  5.64  &  0.10  &  0.02 & --- & --- \\
J1048--0109  &  6.6759  &  5.59  &  0.19  &  0.03 & --- & --- \\
  &   &  5.59  &  0.22  &  0.03 & --- & --- \\
  &   &  5.65  &  0.20  &  0.03 & --- & --- \\
  &   &  5.71  &  0.12  &  0.02 & --- & --- \\
  &   &  5.89  &  0.07  &  0.01 & --- & --- \\
J2002$-$3013  &  6.669  &  5.62  &  0.42  &  0.02 & --- & --- \\
  &   &  5.68  &  0.18  &  0.02 & --- & --- \\
  &   &  5.81  &  0.26  &  0.02 & --- & --- \\
  &   &  5.85  &  0.04  &  0.01 & --- & --- \\
J2232+2930  &  6.666  &  5.64  &  0.08  &  0.01 & --- & --- \\
  &   &  5.69  &  0.06  &  0.01 & --- & --- \\
  &   &  5.69  &  0.02  &  0.004 & --- & --- \\
  &   &  5.70  &  0.01  &  0.003 & --- & --- \\
  &   &  5.70  &  0.02  &  0.005 & --- & --- \\
  &   &  5.70  &  0.05  &  0.01 & --- & --- \\
J1216+4519  &  6.654  &  5.57  &  0.56  &  0.03 & --- & --- \\
  &   &  5.62  &  0.25  &  0.02 & --- & --- \\
  &   &  5.64  &  0.19  &  0.02 & --- & --- \\
  &   &  5.72  &  0.21  &  0.02 & --- & --- \\
  &   &  5.73  &  0.12  &  0.02 & --- & --- \\
  &   &  5.81  &  0.13  &  0.02 & --- & --- \\
J0024+3913  &  6.621  &  5.67  &  0.02  &  0.01 & --- & --- \\
  &   &  5.68  &  0.03  &  0.01 & --- & --- \\
  &   &  5.72  &  0.06  &  0.01 & --- & --- \\
  &   &  5.74  &  0.09  &  0.01 & --- & --- \\
  &   &  5.78  &  0.12  &  0.01 & --- & --- \\
  &   &  5.79  &  0.17  &  0.01 & --- & --- \\
  &   &  5.79  &  0.06  &  0.01 & --- & --- \\
  &   &  6.03  &  0.03  &  0.01  &  0.03  &  0.01 \\
  &   &  6.06  &  0.03  &  0.01  &  0.04  &  0.01 \\
  &   &  6.08  &  0.04  &  0.01  &  0.05  &  0.01 \\
  &   &  6.29  &  0.06  &  0.01  &  0.08  &  0.01 \\
J0305--3150  &  6.6145  &  5.52  &  0.33  &  0.02 & --- & --- \\
  &   &  5.52  &  0.32  &  0.02 & --- & --- \\
  &   &  5.58  &  0.08  &  0.01 & --- & --- \\
  &   &  5.62  &  0.06  &  0.01 & --- & --- \\
  &   &  5.72  &  0.05  &  0.01 & --- & --- \\
  &   &  5.73  &  0.01  &  0.01 & --- & --- \\
  &   &  6.01  &  0.09  &  0.01  &  0.1  &  0.01 \\
  &   &  6.02  &  0.16  &  0.01  &  0.3  &  0.02 \\
  &   &  6.04  &  0.03  &  0.01  &  0.05  &  0.01 \\
J2102--1458  &  6.61  &  5.52  &  0.09  &  0.01 & --- & --- \\
  &   &  5.64  &  0.20  &  0.02 & --- & --- \\
  &   &  5.67  &  0.06  &  0.01 & --- & --- \\
J2132+1217  &  6.5881  &  5.50  &  0.07  &  0.005 & --- & --- \\
  &   &  5.55  &  0.02  &  0.003 & --- & --- \\
  &   &  5.56  &  0.02  &  0.004 & --- & --- \\
  &   &  5.58  &  0.03  &  0.004 & --- & --- \\
  &   &  5.60  &  0.04  &  0.005 & --- & --- \\
  &   &  5.65  &  0.24  &  0.01 & --- & --- \\
  &   &  5.65  &  0.02  &  0.003 & --- & --- \\
  &   &  5.86  &  0.02  &  0.003 & --- & --- \\
  &   &  5.87  &  0.02  &  0.004 & --- & --- \\
  &   &  5.88  &  0.02  &  0.003 & --- & --- \\
  &   &  5.89  &  0.03  &  0.004 & --- & --- \\
J1526--2050  &  6.5864  &  5.54  &  0.07  &  0.01 & --- & --- \\
  &   &  5.55  &  0.02  &  0.004 & --- & --- \\
  &   &  5.56  &  0.01  &  0.003 & --- & --- \\
  &   &  5.56  &  0.07  &  0.01 & --- & --- \\
  &   &  5.58  &  0.30  &  0.01 & --- & --- \\
  &   &  5.58  &  0.40  &  0.01 & --- & --- \\
  &   &  5.59  &  0.11  &  0.01 & --- & --- \\
  &   &  5.60  &  0.14  &  0.01 & --- & --- \\
  &   &  5.61  &  0.48  &  0.01 & --- & --- \\
  &   &  5.62  &  0.36  &  0.01 & --- & --- \\
  &   &  5.63  &  0.01  &  0.003 & --- & --- \\
  &   &  5.70  &  0.05  &  0.005 & --- & --- \\
  &   &  5.71  &  0.04  &  0.004 & --- & --- \\
  &   &  5.72  &  0.09  &  0.005 & --- & --- \\
  &   &  5.73  &  0.17  &  0.01 & --- & --- \\
  &   &  5.75  &  0.18  &  0.01 & --- & --- \\
  &   &  5.76  &  0.17  &  0.01 & --- & --- \\
  &   &  5.83  &  0.13  &  0.01 & --- & --- \\
  &   &  5.83  &  0.21  &  0.01 & --- & --- \\
  &   &  5.84  &  0.08  &  0.01 & --- & --- \\
  &   &  5.84  &  0.04  &  0.005 & --- & --- \\
  &   &  5.86  &  0.15  &  0.01 & --- & --- \\
  &   &  5.96  &  0.10  &  0.01  &  0.11  &  0.01 \\
  &   &  5.97  &  0.05  &  0.01  &  0.06  &  0.01 \\
J1135+5011  &  6.58  &  5.55  &  0.28  &  0.02 & --- & --- \\
  &   &  5.59  &  0.09  &  0.02 & --- & --- \\
  &   &  5.60  &  0.09  &  0.02 & --- & --- \\
  &   &  5.62  &  0.08  &  0.02 & --- & --- \\
  &   &  5.62  &  0.09  &  0.02 & --- & --- \\
  &   &  5.63  &  0.12  &  0.02 & --- & --- \\
  &   &  5.64  &  0.16  &  0.02 & --- & --- \\
  &   &  5.71  &  0.48  &  0.03 & --- & --- \\
  &   &  5.71  &  0.04  &  0.01 & --- & --- \\
  &   &  5.72  &  0.04  &  0.01 & --- & --- \\
  &   &  5.72  &  0.15  &  0.02 & --- & --- \\
  &   &  5.73  &  0.06  &  0.01 & --- & --- \\
  &   &  5.75  &  0.08  &  0.01 & --- & --- \\
  &   &  5.78  &  0.12  &  0.02 & --- & --- \\
  &   &  5.92  &  0.11  &  0.02  &  0.16  &  0.03 \\
  &   &  5.99  &  0.19  &  0.02  &  0.28  &  0.03 \\
  &   &  6.01  &  0.03  &  0.01  &  0.05  &  0.01 \\
  &   &  6.06  &  0.11  &  0.01  &  0.13  &  0.02 \\
J0226+0302  &  6.5412  &  5.51  &  1.11  &  0.01 & --- & --- \\
  &   &  5.53  &  0.35  &  0.01 & --- & --- \\
  &   &  5.54  &  0.05  &  0.004 & --- & --- \\
  &   &  5.55  &  0.04  &  0.003 & --- & --- \\
  &   &  5.55  &  0.02  &  0.003 & --- & --- \\
  &   &  5.55  &  0.01  &  0.002 & --- & --- \\
  &   &  5.56  &  0.03  &  0.003 & --- & --- \\
  &   &  5.56  &  0.01  &  0.003 & --- & --- \\
  &   &  5.57  &  0.03  &  0.003 & --- & --- \\
  &   &  5.57  &  0.12  &  0.01 & --- & --- \\
  &   &  5.59  &  0.02  &  0.003 & --- & --- \\
  &   &  5.60  &  0.08  &  0.01 & --- & --- \\
  &   &  5.61  &  0.05  &  0.004 & --- & --- \\
  &   &  5.62  &  0.07  &  0.004 & --- & --- \\
  &   &  5.63  &  0.16  &  0.01 & --- & --- \\
  &   &  5.65  &  0.04  &  0.004 & --- & --- \\
  &   &  5.67  &  0.06  &  0.005 & --- & --- \\
  &   &  5.69  &  0.01  &  0.002 & --- & --- \\
  &   &  5.69  &  0.05  &  0.004 & --- & --- \\
  &   &  5.70  &  0.05  &  0.003 & --- & --- \\
  &   &  5.71  &  0.03  &  0.003 & --- & --- \\
  &   &  5.73  &  0.02  &  0.003 & --- & --- \\
  &   &  5.84  &  0.13  &  0.005 & --- & --- \\
  &   &  5.85  &  0.01  &  0.002 & --- & --- \\
  &   &  5.86  &  0.02  &  0.003 & --- & --- \\
  &   &  6.29  &  0.01  &  0.003  &  0.02  &  0.01 \\
J0224--4711  &  6.526  &  5.51  &  0.35  &  0.01 & --- & --- \\
  &   &  5.52  &  0.09  &  0.01 & --- & --- \\
  &   &  5.53  &  0.03  &  0.01 & --- & --- \\
  &   &  5.53  &  0.08  &  0.01 & --- & --- \\
  &   &  5.54  &  0.05  &  0.01 & --- & --- \\
  &   &  5.54  &  0.12  &  0.01 & --- & --- \\
  &   &  5.55  &  0.04  &  0.01 & --- & --- \\
  &   &  5.55  &  0.02  &  0.004 & --- & --- \\
  &   &  5.56  &  0.06  &  0.01 & --- & --- \\
  &   &  5.68  &  0.004  &  0.003 & --- & --- \\
  &   &  5.68  &  0.01  &  0.003 & --- & --- \\
  &   &  5.68  &  0.01  &  0.003 & --- & --- \\
  &   &  5.69  &  0.01  &  0.003 & --- & --- \\
  &   &  5.73  &  0.01  &  0.004 & --- & --- \\
J1629+2407  &  6.476  &  5.54  &  0.23  &  0.01 & --- & --- \\
  &   &  5.56  &  0.09  &  0.01 & --- & --- \\
  &   &  5.57  &  0.08  &  0.01 & --- & --- \\
  &   &  5.58  &  0.18  &  0.01 & --- & --- \\
  &   &  5.59  &  0.03  &  0.01 & --- & --- \\
  &   &  5.62  &  0.04  &  0.01 & --- & --- \\
  &   &  5.64  &  0.41  &  0.01 & --- & --- \\
  &   &  5.65  &  0.01  &  0.004 & --- & --- \\
  &   &  5.66  &  0.06  &  0.01 & --- & --- \\
  &   &  5.71  &  0.02  &  0.005 & --- & --- \\
  &   &  5.84  &  0.07  &  0.01 & --- & --- \\
  &   &  5.93  &  0.02  &  0.01  &  0.05  &  0.01 \\
  &   &  5.95  &  0.02  &  0.005  &  0.03  &  0.01 \\
  &   &  5.96  &  0.02  &  0.01  &  0.03  &  0.01 \\
  &   &  6.14  &  0.02  &  0.01  &  0.03  &  0.01 \\
  &   &  6.16  &  0.22  &  0.01  &  0.25  &  0.01 \\
  &   &  6.17  &  0.03  &  0.01  &  0.05  &  0.01 \\
J2318--3113  &  6.4435  &  5.53  &  0.05  &  0.01 & --- & --- \\
  &   &  5.63  &  0.05  &  0.01 & --- & --- \\
  &   &  5.65  &  0.04  &  0.01 & --- & --- \\
  &   &  5.65  &  0.05  &  0.01 & --- & --- \\
  &   &  5.77  &  0.02  &  0.01 & --- & --- \\
  &   &  5.79  &  0.03  &  0.01 & --- & --- \\
  &   &  5.79  &  0.02  &  0.004 & --- & --- \\
  &   &  5.79  &  0.02  &  0.01 & --- & --- \\
  &   &  5.82  &  0.13  &  0.01 & --- & --- \\
  &   &  5.82  &  0.04  &  0.01 & --- & --- \\
  &   &  5.83  &  0.02  &  0.005 & --- & --- \\
J0045+0901  &  6.42  &  5.61  &  0.61  &  0.02 & --- & --- \\
  &   &  5.64  &  0.02  &  0.01 & --- & --- \\
  &   &  5.66  &  0.04  &  0.01 & --- & --- \\
  &   &  5.68  &  0.01  &  0.004 & --- & --- \\
  &   &  5.69  &  0.01  &  0.004 & --- & --- \\
  &   &  5.69  &  0.03  &  0.01 & --- & --- \\
  &   &  5.70  &  0.04  &  0.01 & --- & --- \\
  &   &  5.70  &  0.08  &  0.01 & --- & --- \\
  &   &  5.75  &  0.07  &  0.01 & --- & --- \\
  &   &  5.85  &  0.07  &  0.01 & --- & --- \\
  &   &  6.01  &  0.10  &  0.01  &  0.11  &  0.01 \\
  &   &  6.06  &  0.11  &  0.01  &  0.13  &  0.01 \\
  &   &  6.08  &  0.24  &  0.01  &  0.25  &  0.02 \\
J1148+5251  &  6.4189  &  5.51  &  0.01  &  0.001 & --- & --- \\
  &   &  5.52  &  0.002  &  0.0004 & --- & --- \\
  &   &  5.53  &  0.01  &  0.001 & --- & --- \\
  &   &  5.53  &  0.06  &  0.001 & --- & --- \\
  &   &  5.55  &  0.04  &  0.001 & --- & --- \\
  &   &  5.55  &  0.004  &  0.0005 & --- & --- \\
  &   &  5.56  &  0.03  &  0.001 & --- & --- \\
  &   &  5.57  &  0.02  &  0.001 & --- & --- \\
  &   &  5.59  &  0.02  &  0.001 & --- & --- \\
  &   &  5.59  &  0.06  &  0.001 & --- & --- \\
  &   &  5.60  &  0.02  &  0.001 & --- & --- \\
  &   &  5.61  &  0.002  &  0.0003 & --- & --- \\
  &   &  5.61  &  0.003  &  0.0004 & --- & --- \\
  &   &  5.62  &  0.002  &  0.0003 & --- & --- \\
  &   &  5.62  &  0.01  &  0.0004 & --- & --- \\
  &   &  5.63  &  0.0004  &  0.0002 & --- & --- \\
  &   &  5.63  &  0.03  &  0.001 & --- & --- \\
  &   &  5.64  &  0.04  &  0.001 & --- & --- \\
  &   &  5.65  &  0.03  &  0.001 & --- & --- \\
  &   &  5.66  &  0.12  &  0.001 & --- & --- \\
  &   &  5.67  &  0.01  &  0.0004 & --- & --- \\
  &   &  5.71  &  0.001  &  0.0003 & --- & --- \\
  &   &  5.71  &  0.001  &  0.0003 & --- & --- \\
  &   &  5.72  &  0.002  &  0.0004 & --- & --- \\
  &   &  5.73  &  0.06  &  0.001 & --- & --- \\
  &   &  5.73  &  0.003  &  0.0004 & --- & --- \\
  &   &  5.80  &  0.02  &  0.001 & --- & --- \\
  &   &  5.81  &  0.01  &  0.001 & --- & --- \\
  &   &  5.85  &  0.02  &  0.001 & --- & --- \\
  &   &  5.86  &  0.01  &  0.001 & --- & --- \\
  &   &  5.87  &  0.005  &  0.001 & --- & --- \\
  &   &  5.89  &  0.01  &  0.001 & --- & --- \\
  &   &  5.90  &  0.01  &  0.001  &  0.01  &  0.001 \\
  &   &  5.91  &  0.001  &  0.0003  &  0.003  &  0.001 \\
  &   &  5.91  &  0.01  &  0.001  &  0.01  &  0.001 \\
  &   &  6.08  &  0.02  &  0.001  &  0.02  &  0.001 \\
J1036--0232  &  6.3809  &  5.57  &  0.06  &  0.01 & --- & --- \\
  &   &  5.57  &  0.03  &  0.004 & --- & --- \\
  &   &  5.58  &  0.02  &  0.004 & --- & --- \\
  &   &  5.59  &  0.10  &  0.01 & --- & --- \\
  &   &  5.60  &  0.02  &  0.004 & --- & --- \\
  &   &  5.62  &  0.01  &  0.003 & --- & --- \\
  &   &  5.63  &  0.03  &  0.004 & --- & --- \\
  &   &  5.92  &  0.03  &  0.005  &  0.04  &  0.01 \\
  &   &  5.94  &  0.04  &  0.005  &  0.04  &  0.01 \\
  &   &  5.95  &  0.03  &  0.005  &  0.04  &  0.01 \\
  &   &  5.96  &  0.16  &  0.01  &  0.17  &  0.01 \\
J1152+0055  &  6.3637  &  5.50  &  0.13  &  0.01 & --- & --- \\
  &   &  5.53  &  0.02  &  0.01 & --- & --- \\
  &   &  5.53  &  0.05  &  0.01 & --- & --- \\
  &   &  5.66  &  0.20  &  0.02 & --- & --- \\
  &   &  5.66  &  0.04  &  0.01 & --- & --- \\
J1148+0702  &  6.339  &  5.52  &  0.04  &  0.01 & --- & --- \\
  &   &  5.55  &  0.05  &  0.01 & --- & --- \\
  &   &  5.56  &  0.06  &  0.01 & --- & --- \\
  &   &  5.58  &  0.20  &  0.01 & --- & --- \\
  &   &  5.59  &  0.04  &  0.01 & --- & --- \\
  &   &  5.62  &  0.04  &  0.01 & --- & --- \\
  &   &  5.62  &  0.05  &  0.01 & --- & --- \\
  &   &  5.63  &  0.14  &  0.01 & --- & --- \\
  &   &  5.63  &  0.04  &  0.01 & --- & --- \\
  &   &  5.65  &  0.02  &  0.004 & --- & --- \\
  &   &  5.66  &  0.02  &  0.004 & --- & --- \\
  &   &  5.67  &  0.03  &  0.005 & --- & --- \\
  &   &  5.69  &  0.05  &  0.004 & --- & --- \\
  &   &  5.69  &  0.08  &  0.01 & --- & --- \\
  &   &  5.70  &  0.15  &  0.01 & --- & --- \\
  &   &  5.71  &  0.02  &  0.003 & --- & --- \\
  &   &  5.75  &  0.02  &  0.004 & --- & --- \\
  &   &  5.76  &  0.02  &  0.004 & --- & --- \\
  &   &  5.79  &  0.09  &  0.01 & --- & --- \\
  &   &  5.82  &  0.09  &  0.01 & --- & --- \\
  &   &  5.83  &  0.05  &  0.01 & --- & --- \\
  &   &  5.89  &  0.11  &  0.01 & --- & --- \\
  &   &  5.92  &  0.03  &  0.01  &  0.03  &  0.01 \\
J0142--3327  &  6.3379  &  5.58  &  0.08  &  0.005 & --- & --- \\
  &   &  5.58  &  0.05  &  0.01 & --- & --- \\
  &   &  5.59  &  0.33  &  0.02 & --- & --- \\
  &   &  5.64  &  0.06  &  0.01 & --- & --- \\
  &   &  5.68  &  0.04  &  0.01 & --- & --- \\
  &   &  5.68  &  0.02  &  0.004 & --- & --- \\
  &   &  5.68  &  0.01  &  0.004 & --- & --- \\
  &   &  5.69  &  0.02  &  0.005 & --- & --- \\
  &   &  5.69  &  0.03  &  0.005 & --- & --- \\
  &   &  5.76  &  0.05  &  0.01 & --- & --- \\
  &   &  5.78  &  0.29  &  0.01 & --- & --- \\
  &   &  5.79  &  0.02  &  0.004 & --- & --- \\
  &   &  5.79  &  0.04  &  0.005 & --- & --- \\
  &   &  5.79  &  0.12  &  0.01 & --- & --- \\
  &   &  5.80  &  0.02  &  0.004 & --- & --- \\
J0100+2802  &  6.327  &  5.51  &  0.01  &  0.0004 & --- & --- \\
  &   &  5.52  &  0.003  &  0.0003 & --- & --- \\
  &   &  5.54  &  0.71  &  0.003 & --- & --- \\
  &   &  5.55  &  0.002  &  0.0002 & --- & --- \\
  &   &  5.56  &  0.04  &  0.001 & --- & --- \\
  &   &  5.56  &  0.002  &  0.0002 & --- & --- \\
  &   &  5.56  &  0.002  &  0.0002 & --- & --- \\
  &   &  5.56  &  0.02  &  0.001 & --- & --- \\
  &   &  5.57  &  0.01  &  0.0003 & --- & --- \\
  &   &  5.57  &  0.06  &  0.001 & --- & --- \\
  &   &  5.58  &  0.02  &  0.0004 & --- & --- \\
  &   &  5.59  &  0.19  &  0.002 & --- & --- \\
  &   &  5.61  &  0.05  &  0.001 & --- & --- \\
  &   &  5.62  &  0.10  &  0.001 & --- & --- \\
  &   &  5.63  &  0.002  &  0.0002 & --- & --- \\
  &   &  5.63  &  0.07  &  0.001 & --- & --- \\
  &   &  5.64  &  0.03  &  0.0004 & --- & --- \\
  &   &  5.65  &  0.005  &  0.0002 & --- & --- \\
  &   &  5.65  &  0.01  &  0.0003 & --- & --- \\
  &   &  5.65  &  0.002  &  0.0002 & --- & --- \\
  &   &  5.67  &  0.03  &  0.0004 & --- & --- \\
  &   &  5.67  &  0.01  &  0.0003 & --- & --- \\
  &   &  5.68  &  0.06  &  0.001 & --- & --- \\
  &   &  5.68  &  0.003  &  0.0002 & --- & --- \\
  &   &  5.68  &  0.02  &  0.0003 & --- & --- \\
  &   &  5.69  &  0.005  &  0.0002 & --- & --- \\
  &   &  5.69  &  0.004  &  0.0002 & --- & --- \\
  &   &  5.70  &  0.003  &  0.0002 & --- & --- \\
  &   &  5.75  &  0.001  &  0.0002 & --- & --- \\
  &   &  5.76  &  0.0004  &  0.0001 & --- & --- \\
  &   &  5.78  &  0.004  &  0.0002 & --- & --- \\
  &   &  5.84  &  0.02  &  0.0004 & --- & --- \\
  &   &  5.87  &  0.004  &  0.0003 & --- & --- \\
  &   &  5.88  &  0.04  &  0.001 & --- & --- \\
  &   &  5.89  &  0.0004  &  0.0001 & --- & --- \\
  &   &  5.89  &  0.001  &  0.0001 & --- & --- \\
  &   &  5.96  &  0.0003  &  0.0001  &  0.001  &  0.0002 \\
  &   &  5.96  &  0.001  &  0.0002  &  0.002  &  0.001 \\
  &   &  5.97  &  0.002  &  0.0002  &  0.002  &  0.0004 \\
  &   &  5.99  &  0.002  &  0.0002  &  0.002  &  0.0003 \\
  &   &  5.99  &  0.09  &  0.001  &  0.09  &  0.001 \\
  &   &  6.02  &  0.002  &  0.0002  &  0.002  &  0.0002 \\
J1030+0524  &  6.308  &  5.51  &  0.03  &  0.002 & --- & --- \\
  &   &  5.52  &  0.02  &  0.002 & --- & --- \\
  &   &  5.52  &  0.02  &  0.002 & --- & --- \\
  &   &  5.53  &  0.01  &  0.001 & --- & --- \\
  &   &  5.54  &  0.22  &  0.003 & --- & --- \\
  &   &  5.56  &  0.02  &  0.002 & --- & --- \\
  &   &  5.56  &  0.003  &  0.001 & --- & --- \\
  &   &  5.57  &  0.38  &  0.01 & --- & --- \\
  &   &  5.58  &  0.15  &  0.003 & --- & --- \\
  &   &  5.59  &  0.04  &  0.002 & --- & --- \\
  &   &  5.59  &  0.15  &  0.003 & --- & --- \\
  &   &  5.61  &  0.01  &  0.001 & --- & --- \\
  &   &  5.62  &  0.04  &  0.002 & --- & --- \\
  &   &  5.62  &  0.09  &  0.003 & --- & --- \\
  &   &  5.63  &  0.04  &  0.003 & --- & --- \\
  &   &  5.63  &  0.02  &  0.002 & --- & --- \\
  &   &  5.64  &  0.02  &  0.001 & --- & --- \\
  &   &  5.65  &  0.003  &  0.001 & --- & --- \\
  &   &  5.65  &  0.02  &  0.002 & --- & --- \\
  &   &  5.66  &  0.10  &  0.002 & --- & --- \\
  &   &  5.67  &  0.02  &  0.001 & --- & --- \\
  &   &  5.67  &  0.02  &  0.002 & --- & --- \\
  &   &  5.68  &  0.01  &  0.001 & --- & --- \\
  &   &  5.69  &  0.01  &  0.001 & --- & --- \\
  &   &  5.69  &  0.01  &  0.001 & --- & --- \\
  &   &  5.69  &  0.04  &  0.002 & --- & --- \\
  &   &  5.70  &  0.01  &  0.001 & --- & --- \\
  &   &  5.70  &  0.003  &  0.001 & --- & --- \\
  &   &  5.71  &  0.06  &  0.002 & --- & --- \\
  &   &  5.71  &  0.005  &  0.001 & --- & --- \\
  &   &  5.72  &  0.01  &  0.001 & --- & --- \\
  &   &  5.73  &  0.01  &  0.001 & --- & --- \\
  &   &  5.76  &  0.01  &  0.001 & --- & --- \\
  &   &  5.77  &  0.02  &  0.002 & --- & --- \\
  &   &  5.78  &  0.02  &  0.002 & --- & --- \\
  &   &  5.78  &  0.03  &  0.002 & --- & --- \\
  &   &  5.79  &  0.002  &  0.001 & --- & --- \\
  &   &  5.79  &  0.06  &  0.002 & --- & --- \\
  &   &  5.80  &  0.05  &  0.002 & --- & --- \\
  &   &  5.80  &  0.004  &  0.001 & --- & --- \\
  &   &  5.81  &  0.03  &  0.002 & --- & --- \\
  &   &  5.84  &  0.003  &  0.001 & --- & --- \\
  &   &  5.84  &  0.03  &  0.002 & --- & --- \\
  &   &  5.85  &  0.01  &  0.001 & --- & --- \\
  &   &  5.86  &  0.13  &  0.002 & --- & --- \\
  &   &  5.86  &  0.03  &  0.002 & --- & --- \\
  &   &  5.90  &  0.02  &  0.002 & --- & --- \\
  &   &  5.91  &  0.003  &  0.001  &  0.004  &  0.001 \\
  &   &  5.92  &  0.01  &  0.001  &  0.01  &  0.002 \\
  &   &  5.95  &  0.03  &  0.001  &  0.03  &  0.002 \\
\enddata
\label{tab:spike-alpha}
\end{deluxetable}

\startlongtable
\begin{deluxetable}{l l l l l l l}
\tablecaption{Ly$\beta$ Transmission Spikes Identified From Our Quasar Sample. }
\tablewidth{\textwidth}
\tablehead{
\colhead{Object} &
\colhead{$z_{\rm em}$} &
\colhead{$z_{\rm spike}$} &
\colhead{$W$} &
\colhead{$\sigma_{W}$} &
\colhead{EW} &
\colhead{$\sigma_{\rm EW}$} 
}
\startdata
J0024+3913  &  6.621  &  6.26  &  0.04  &  0.01  &  0.05  &  0.01 \\
  &   &  6.29  &  0.05  &  0.01  &  0.06  &  0.01 \\
J0305-3150  &  6.6145  &  6.28  &  0.08  &  0.01  &  0.08  &  0.01 \\
J2132+1217  &  6.5881  &  6.22  &  0.11  &  0.01  &  0.11  &  0.01 \\
  &   &  6.23  &  0.04  &  0.004  &  0.05  &  0.01 \\
  &   &  6.25  &  0.05  &  0.01  &  0.06  &  0.01 \\
  &   &  6.26  &  0.09  &  0.01  &  0.09  &  0.01 \\
  &   &  6.30  &  0.03  &  0.004  &  0.04  &  0.01 \\
J1526-2050  &  6.5864  &  6.24  &  0.03  &  0.003  &  0.05  &  0.005 \\
J1135+5011  &  6.58  &  6.27  &  0.09  &  0.01  &  0.11  &  0.02 \\
J0226+0302  &  6.5412  &  6.29  &  0.11  &  0.01  &  0.11  &  0.01 \\
J0224-4711  &  6.526  &  6.20  &  0.03  &  0.005  &  0.04  &  0.01 \\
  &   &  6.21  &  0.01  &  0.004  &  0.02  &  0.01 \\
  &   &  6.21  &  0.07  &  0.01  &  0.07  &  0.01 \\
  &   &  6.27  &  0.01  &  0.003  &  0.01  &  0.01 \\
J2318-3113  &  6.4435  &  6.15  &  0.05  &  0.01  &  0.07  &  0.01 \\
J1148+5251  &  6.4189  &  6.06  &  0.01  &  0.0003  &  0.01  &  0.0003 \\
  &   &  6.09  &  0.002  &  0.0002  &  0.002  &  0.0002 \\
  &   &  6.15  &  0.01  &  0.001  &  0.01  &  0.001 \\
  &   &  6.16  &  0.001  &  0.0004  &  0.003  &  0.001 \\
J1036-0232  &  6.3809  &  6.03  &  0.02  &  0.004  &  0.03  &  0.01 \\
  &   &  6.07  &  0.03  &  0.005  &  0.04  &  0.01 \\
  &   &  6.08  &  0.02  &  0.005  &  0.03  &  0.01 \\
  &   &  6.09  &  0.11  &  0.01  &  0.11  &  0.01 \\
  &   &  6.10  &  0.01  &  0.003  &  0.01  &  0.01 \\
J1152+0055  &  6.3637  &  6.10  &  0.04  &  0.01  &  0.06  &  0.02 \\
  &   &  6.11  &  0.03  &  0.01  &  0.05  &  0.01 \\
J1148+0702  &  6.339  &  6.00  &  0.03  &  0.004  &  0.03  &  0.01 \\
  &   &  6.04  &  0.01  &  0.002  &  0.02  &  0.01 \\
  &   &  6.04  &  0.05  &  0.01  &  0.05  &  0.01 \\
  &   &  6.10  &  0.02  &  0.004  &  0.03  &  0.01 \\
J0142-3327  &  6.3379  &  5.99  &  0.02  &  0.005  &  0.03  &  0.01 \\
  &   &  6.00  &  0.01  &  0.004  &  0.01  &  0.01 \\
  &   &  6.05  &  0.005  &  0.003  &  0.01  &  0.01 \\
  &   &  6.06  &  0.09  &  0.01  &  0.09  &  0.01 \\
  &   &  6.07  &  0.02  &  0.004  &  0.03  &  0.01 \\
J0100+2802  &  6.3258  &  5.97  &  0.01  &  0.001  &  0.01  &  0.0004 \\
  &   &  5.98  &  0.04  &  0.001  &  0.04  &  0.0005 \\
  &   &  5.99  &  0.27  &  0.01  &  0.22  &  0.001 \\
  &   &  6.01  &  0.001  &  0.0001  &  0.001  &  0.0002 \\
  &   &  6.02  &  0.001  &  0.0002  &  0.002  &  0.0003 \\
  &   &  6.02  &  0.02  &  0.001  &  0.02  &  0.0004 \\
  &   &  6.03  &  0.01  &  0.001  &  0.02  &  0.0003 \\
  &   &  6.04  &  0.002  &  0.0002  &  0.002  &  0.0003 \\
J1030+0524  &  6.308  &  5.95  &  0.02  &  0.001  &  0.02  &  0.002 \\
  &   &  5.96  &  0.004  &  0.001  &  0.01  &  0.001 \\
  &   &  5.96  &  0.001  &  0.001  &  0.003  &  0.001 \\
  &   &  5.97  &  0.01  &  0.001  &  0.01  &  0.002 \\
  &   &  5.97  &  0.01  &  0.001  &  0.01  &  0.002 \\
  &   &  5.97  &  0.01  &  0.001  &  0.01  &  0.002 \\
\enddata
\label{tab:spike-beta}
\end{deluxetable}

\begin{figure*}
\centering 
\includegraphics[width=1.0\textwidth]{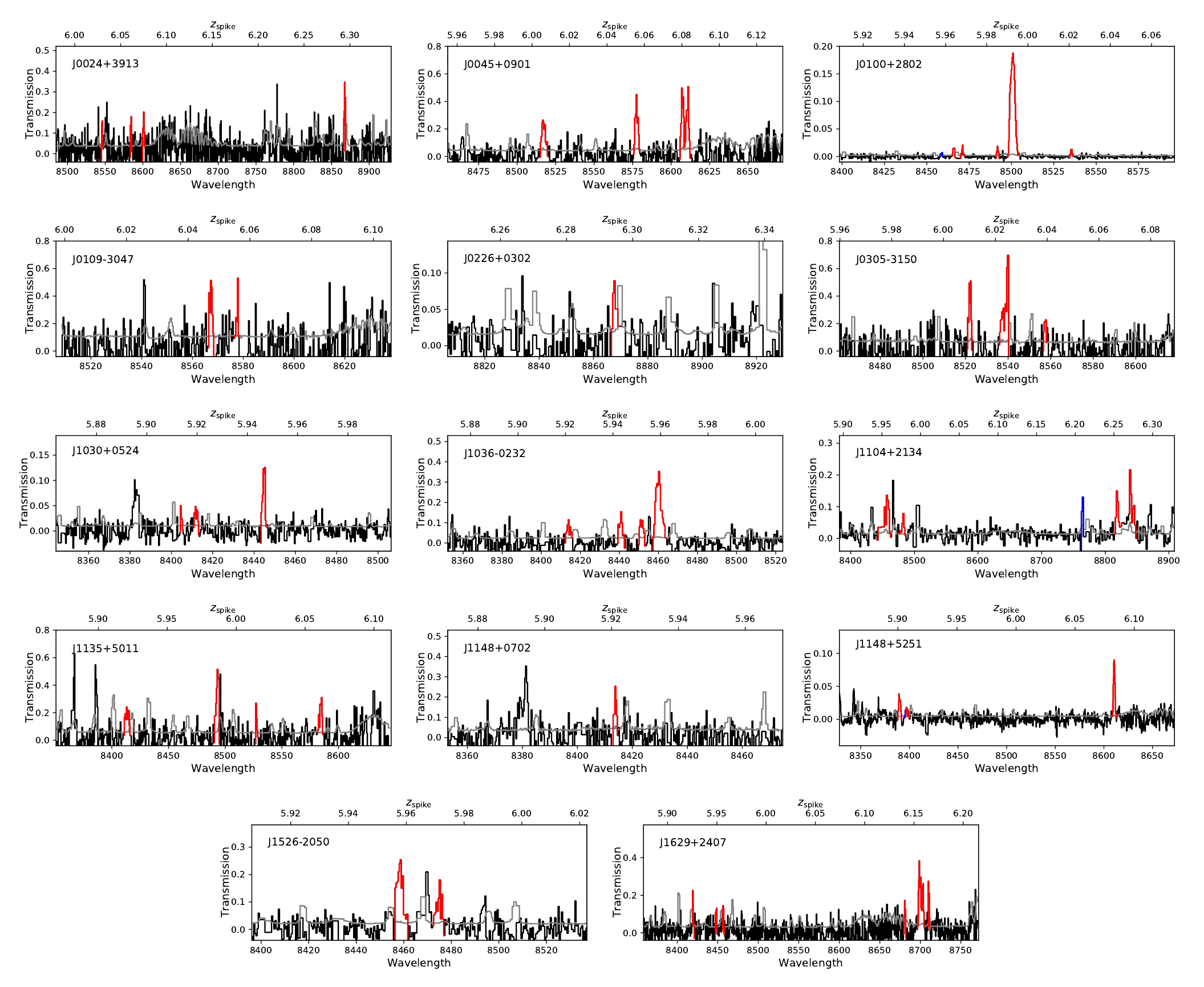}
\caption{Ly$\alpha$ transmission spikes at $z > 5.9$ identified in our sample of sightlines within the Ly$\alpha$ forest windows. They are all shown in both 1D spectra and 2D images. The spikes with $> 3 \sigma$ detection are in red, while the $< 3 \sigma$ spikes are shown in blue.}
\label{fig:spike-alpha}
\end{figure*}

\begin{figure*}
\centering 
\includegraphics[width=1.0\textwidth]{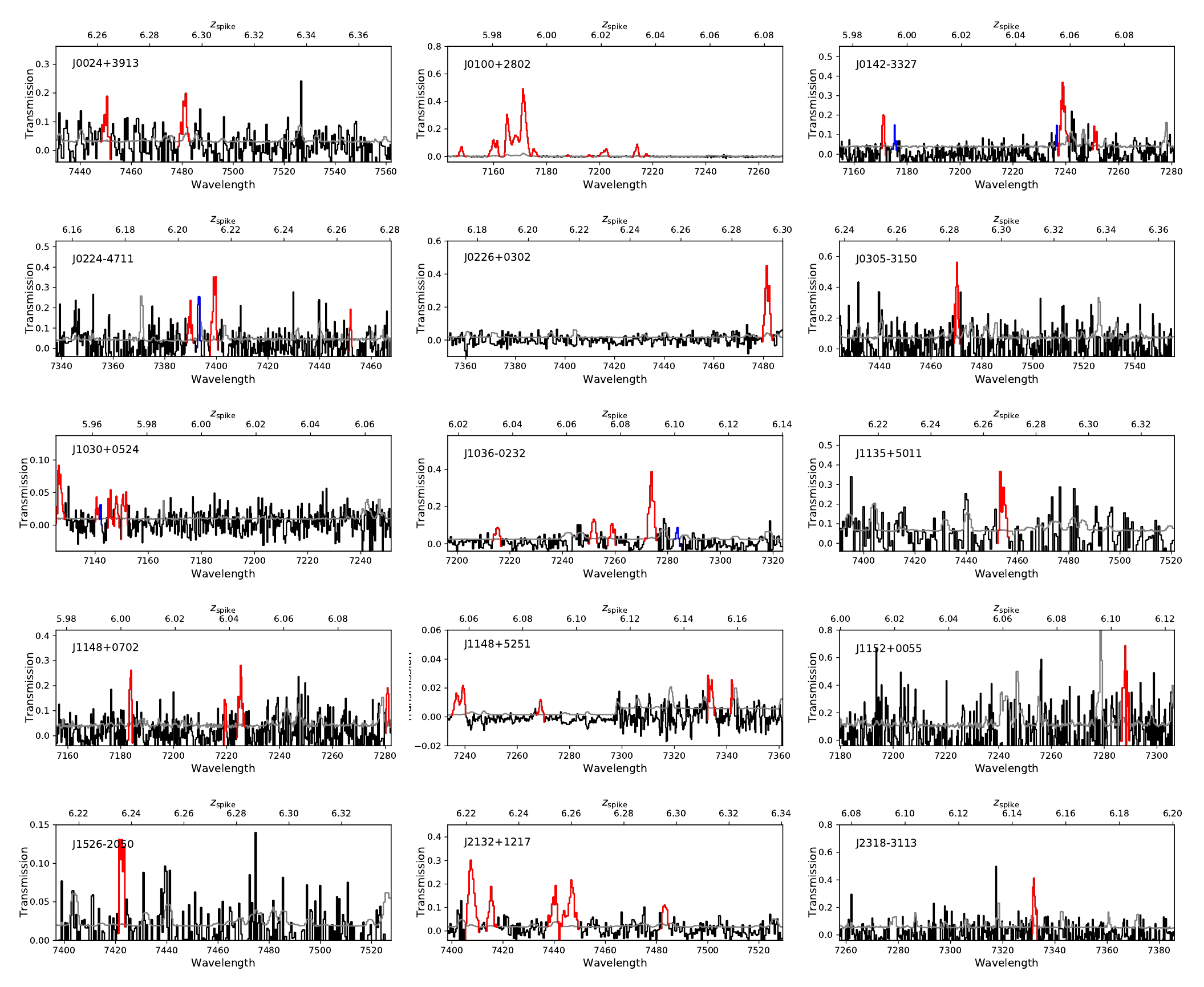}
\caption{Ly$\beta$ transmission spikes at $z > 5.9$ identified in our sample of sightlines within the Ly$\beta$ forest windows. They are all shown in both 1D spectra and 2D images. The spikes with $> 3 \sigma$ detection are in red, while the $< 3 \sigma$ spikes are in blue.}
\label{fig:spike-beta}
\end{figure*}

\end{document}